\documentclass[10pt,journal,compsoc]{IEEEtran}

\usepackage[numbers]{natbib}
\RequirePackage[colorlinks,citecolor=blue,urlcolor=blue]{hyperref}
\usepackage{placeins}
\usepackage[caption=false]{subfig}
\ifCLASSINFOpdf
  \usepackage[pdftex]{graphicx}
\else
  \usepackage[dvips]{graphicx}
\fi
\usepackage{amsmath}
\begin{document}
%
\title{Detection of cybersecurity attacks through analysis of web browsing activities using principal component analysis}
%
%
%
%

\author{Insha~Ullah,
        Kerrie~Mengersen,
        Rob~J~Hyndman,
        and~James~McGree
\IEEEcompsocitemizethanks{\IEEEcompsocthanksitem Insha Ullah was a research associate in the Science and Engineering Faculty, Queensland University of Technology,
QLD, Australia, 4000.\protect\\
E-mail: i.ullah1980@gmail.com
\IEEEcompsocthanksitem Kerrie Mengersen and James McGree are with the Queensland University of Technology, Australia.
\IEEEcompsocthanksitem Rob J Hyndman is with Monash University, Australia.
}
}

\IEEEtitleabstractindextext{%
\begin{abstract}
Organizations such as government departments and financial institutions provide online service facilities accessible via an increasing number of internet connected devices which make their operational environment vulnerable to cyber attacks.
Consequently, there is a need to have mechanisms in place to detect cyber security attacks in a timely manner. A variety of Network Intrusion Detection Systems (NIDS) have been proposed and can be categorized into signature-based NIDS and anomaly-based NIDS.
The signature-based NIDS, which identify the misuse through scanning the activity signature against the list of known attack activities, are criticized for their inability to identify new attacks (never-before-seen attacks).
Among anomaly-based NIDS, which declare a connection anomalous if it expresses deviation from a trained model, the unsupervised learning algorithms circumvent this issue since they have the ability to identify new attacks.
In this study, we use an unsupervised learning algorithm based on principal component analysis to detect cyber attacks. In the training phase, our approach has the advantage of also identifying outliers in the training dataset. In the monitoring phase, our approach first identifies the affected dimensions and then calculates an anomaly score by aggregating across only those components that are affected by the anomalies. We explore the performance of the algorithm via simulations and through two applications, namely to the UNSW-NB15 dataset recently released by the Australian Centre for Cyber Security and to the well-known KDD'99 dataset.
The algorithm is scalable to large datasets in both training and monitoring phases, and the results from both the simulated and real datasets show that the method has promise in detecting suspicious network activities.
\end{abstract}

}

\maketitle

\IEEEdisplaynontitleabstractindextext

%
\IEEEpeerreviewmaketitle

\ifCLASSOPTIONcompsoc
\IEEEraisesectionheading{\section{Introduction}\label{sec:introduction}}
\else
\section{Introduction}
\label{sec:introduction}
\fi

%
%
%
%
The vulnerability of government departments and industries to cyber attacks is rising due to the increased usage and reliance on network-connected platforms and internet-connected devices. As such, there is a need to not only have secure cyber systems but also have mechanisms in place to detect cyber security attacks in a timely manner.
The analysis of automatically generated logs from network and system devices can provide deep insight into the current state of security of computer systems. In addition, effective threat intelligence can be performed which can proactively predict several possible attack vectors relevant to government departments. However, there are particular challenges in dealing with such data, including the sheer volume of events of web browsing activity. Thus, computationally efficient approaches are needed for handling and analyzing such data.

Network intrusion detection, as a particular form of anomaly detection, is an active area of research.
Many Network Intrusion Detection Systems (NIDS) are designed to protect against malicious activities on the network \citep[e.g. see][for references]{bhuyan2014network, buczak2016survey,sahoo2017malicious}. These systems can be classified into two main categories \citep{shyu2005handling}: signature-based NIDS (also known as misuse-based NIDS) and anomaly-based NIDS \citep{garcia2009anomaly,om2012statistical}. In signature-based NIDS, misuse is identified by scanning the activities' signatures against a database of known malicious activities signatures. Many companies (such as Splunk\footnote{\url{https://www.splunk.com}} and Sumo Logic\footnote{\url{https://www.sumologic.com}}) and open source packages (such as ElasticSearch\footnote{\url{https://www.elastic.co/}} and Graylog\footnote{\url{https://www.graylog.org/}}) provide log analysis services based on static signature matching. Although the signature-based NIDS have seen high accuracies and low false alarm rates against the known type of attacks, they lack the ability to identify new attacks due to the incomplete list of known malicious activities. Another shortcoming of the signature-based NIDS is that the attackers need to make only a slight modification to their signature to avoid them \citep{prakash2010phishnet}. On the other hand, in an anomaly-based NIDS a profile of normal activities is built using machine learning algorithms and future activities are scanned against the normal profile.

A variety of supervised and unsupervised statistical machine learning algorithms have been proposed for intrusion detection. In general, the supervised learning (classification) algorithms are known to have superior performance compared with the unsupervised (clustering) algorithms in terms of classification accuracy. However, their requirement of having a pre-labelled training dataset is often unavailable in many applications. For example, large networks can easily produce hundreds of gigabytes in logs each day. A large training dataset may be required to learn the normal behavior of such a large system. Further, the normal activities patterns may change over time, which may require retraining of the previously fitted model periodically. In such cases, it may not always be feasible to label each instance of a network activities dataset as normal or malicious since it requires human involvement to review and classify training data, and this requirement does not scale to large data sets. Further, some anomalies may be unknown or at least unseen in the training dataset.

Consequently, unsupervised learning methods seem more appealing for monitoring network activities since they do not require a pre-labelled training dataset. However, they do require a clean dataset (free of attacks) if the monitoring objective is to uncover both known and new types of attacks but we note that this is generally easy to acquire \citep{ringberg2007sensitivity}. Precisely, an unsupervised anomaly-based NIDS works in two phases: training and monitoring. In the training phase, a model or an algorithm is used to learn the normal behavior of a system using a `normal' activity dataset. In the monitoring phase, the learned model is used to evaluate the behavior of the future network traffic. A behavior that is inconsistent with the trained model triggers an alert such that the activity can be further investigated. If the investigation leads to a declaration of the identified anomaly as normal activity, this activity should be included in the normal profile and the model is re-trained for future scanning. Unsupervised anomaly-based NIDS have the ability to identify what is referred to as zero-day or never-before-seen attacks \citep{perdisci2006using}.
Note also that in some applications the objective may be to detect only new attacks, in which case the training dataset is allowed to contain the known types of attacks.

A variety of unsupervised learning algorithms have already been adapted for intrusion detection. However, some of these are computationally intensive and require large computer memory to operate in practice \citep[e.g. see][]{muda2011intrusion, horng2011novel}. Dimension reduction approaches such as principal component analysis (PCA) have also been proposed \citep{shyu2003novel,lakhina2004characterization}. These are appealing because they are statistically coherent, computationally faster and scalable. However, PCA-based NIDS have been criticized by \cite{ringberg2007sensitivity} for several reasons. First, these NIDS are sensitive to the choice of the number of principal components that capture the normal activities patterns. Often ad~hoc approaches are used to make this choice. Second, the effectiveness of PCA is sensitive to the level of aggregation of the traffic measurements. Third, the presence of unusual observations in the normal activities data that are used for training purpose makes the NIDS less effective. Fourth, correctly identifying which flow triggered the anomaly detector is an inherently challenging problem.

Most of the PCA-based NIDS developed so far share disadvantages highlighted by \cite{ringberg2007sensitivity}. For example, \cite{wang2006identifying} and \cite{callegari2011novel} have used the proportion of variance captured and the scree-plot method, respectively, to choose the number of components. These ad~hoc procedures to select the number of PCs render the PCA less effective, since it is likely to miss the components that describe the anomaly if they are among the minor components (last PCs that explain less variance and are often dropped from the analysis). To address the third issue; that is, the existence of unusual observations in the normal activities data, \cite{kwitt2007unsupervised} have used PCA based on robust estimators; however, the selection of the number of components is still an unsolved problem in their study.

A large number of variables could be recorded to describe the behavior of the network traffic \citep{lee2000framework}. It could be that an anomaly is different from normal activities patterns with respect to fewer variables and the other variables are irrelevant. This is assumed by \cite{koeman2019critical}, for example, and is the usual assumption of variable selection methods. It may also be that a set of variables that distinguishes one kind of anomaly from normal activities behavior differs from the set that are useful in identifying another kind of anomaly. PCA seeks to derive new variables called principal components (PCs) as linear combinations of the input variables such that a few of the new variables account for the maximum variability among the input variables. If the data are similarly scaled, the magnitude of the coefficients (also referred to as loadings) in a PC reflect relative importance of the respective input variables in describing the variation ascribed to that PC and the signs of the coefficients specify positive or negative correlation between the input variables. Typically each input variable has a non-zero coefficient across all PCs even if the variables are just a random noise. However, in practice it is common that majority of the component loadings are close to zero \citep{zou2006sparse} and are of less practical significance. A variable that makes an anomaly detectable in a large number of observations is important and a PC that has a large loading for this important variable needs to be taken into account when calculating scores for anomaly detection. It is also likely that this important variable has high correlation with all other variables that too are useful in identifying the same anomaly, hence are likely to have larger loadings for the same PCs. 

In this paper we propose to use PCA to more effectively detect unusual web browsing activities and attempt to address the issues raised by \cite{ringberg2007sensitivity}. In the training phase we fit the PCA model using the normal activities dataset and find the bootstrap distribution of the standard deviations of the standardized PCs. These bootstrap distributions are useful in identifying the important PCs and also helps in locating large influential observations if there are any in the training set of data. In the monitoring phase, we take into account fewer important components to calculate anomaly scores for future connections. The scores of anomalous activities based on only important components are expected to deviate substantially from the scores obtained for normal connections and hence should be useful in identifying intrusions.
For illustration we consider the performance of this approach in a simulation study and through application to two publicly available datasets including UNSW-NB15 dataset recently released by the Australian Centre for Cyber Security.
Our method is straightforward to understand, computationally efficient to implement and empirical evaluation on several real-world datasets suggests that it is more effective in detecting abnormal activities when compared to existing methods.

\section{Principal component analysis}
\label{section2}
PCA is traditionally used as a dimension reduction technique since it tends to capture as much variation as possible in a few components and is a useful tool to visualize high-dimensional data in a manageable low-dimensional space \citep{jolliffe2002principal,alter2000singular,mctavish2013new}. It uses an orthogonal transformation to transform a set of $p$ (possibly correlated) observed variables into another set of $p$ uncorrelated variables. The new derived set of variables are the $p$ principal components (PCs). Each PC is a linear combination of the originally observed variables such that the first PC stands for the axis along which the observed data exhibit the highest variance, the second PC stands for the axis that is orthogonal to the first PC and along which the observed data exhibit the second highest variance, the third PC stands for the axis that is orthogonal to the first two PCs and along which the observed data exhibit the third highest variance, and so on. Similarly, the $p$th PC (the last one) stands for the axis that is orthogonal to the remaining $p-1$ PCs and along which the observed data exhibit the least variance. In this way, the $p$ orthogonal dimensions of variability that exist in the data are captured in $p$ PCs and the proportion of variability that each PC accounts for accumulates to the total variation in the data. It is often the case that the first $q$ ($q \ll p$) PCs retain interpretable patterns in the observed data and the rest contain variation mainly due to noise \citep{jolliffe2002principal}.

More formally, let $y_{ij}$ denote a real-valued observation made of the $j$th variable on the $i$th subject, where $i=1,\hdots,n$ and $j=1,\hdots,p$. Assume that the $n$ observations are arranged in $n$ rows of a $n\times p$ matrix ${Y}$ with columns corresponding to $p$ variables or features. We standardize columns of matrix $Y$ to have zero mean and unit standard deviation, and store the resultant values in a matrix $X$, that is, the elements $x_{ij}$ of $X$ are obtained by $x_{ij}=(y_{ij}-\bar{y}_j)/{s_j}$, where $\bar{y}_j$ and $s_j$ are the mean and standard deviation of the $j$th column of matrix $Y$, respectively.
The PCA can be performed by singular value decomposition (SVD) of $X$, that is, the $n\times p$ matrix is decomposed as
\begin{equation}
X=U\Gamma V^{T},
\label{svd}
\end{equation}
where $U$ is $n\times p$ matrix with orthonormal columns, $\Gamma$ is a $p\times p$ diagonal matrix containing non-negative singular values and $V$ is a $p\times p$ matrix with orthonormal columns. Denote the sample correlation matrix of $X$ by $R$, which can be expressed as
\begin{equation}
R = \frac{1}{n-1}X^TX
  = V\Lambda V^T,
\label{eigds}
\end{equation}
where $\Lambda=1/(n-1)\Gamma^2$ is a $p\times p$ diagonal matrix containing $p$ eigenvalues $\boldsymbol{\lambda}=(\lambda_1,\hdots,\lambda_p)^T$ on the diagonal in decreasing order of magnitude. It follows that the $p$ columns of matrix $V$ contain the eigenvectors of $R$ and hence are the desired axes of variation. The derived set of $p$ transformed variables (the PCs) are obtained by
$$
Z = XV.
$$
It is important to note that the matrix $U$ above contains standardized PCs in columns and is a scaled version of $Z$, which is provided additionally in \eqref{svd}. To see this, multiply \eqref{svd} on the right by $V$ as follows
\begin{equation*}
XV=U\Gamma
\end{equation*}
or
\begin{equation*}
Z=U\Gamma.
\end{equation*}
In practice, the first $q$ major components are of greater interest since they account for most of the variation in the data and the dimension of $Z$ is thus reduced from $p$ to $q$, that is
$$
\tilde{Z} = X\tilde{V},
$$
where $\tilde{V}$ is a $p\times q$ matrix that contains the first $q$ columns of $V$ and $\tilde{Z}$ contains the first $q$ PCs. A lower rank approximation of $X$ can be obtained by
\begin{equation}
\tilde{X} = \tilde{Z}\tilde{V}^T,
\label{appx}
\end{equation}
which is the best approximation of $X$ in the least-square-sense by a matrix of rank $q$ \citep{saporta2009principal}.

If $X$ has a multivariate normal distribution, then the statistic
\begin{equation}
t_i = \sum_{j=1}^q \frac{z_{ij}^2}{\lambda_j}=(n-1)\sum_{j=1}^q u_{ij}^2,
\label{ti}
\end{equation}
follows a chi-square distribution with $q$ degrees of freedom that is $t_i\sim\chi^2_q$ distribution \citep{shyu2003novel, saporta2009principal}, where $z_{ij}$ and $u_{ij}$ are the elements of $Z$ and $U$.
Determining a suitable value of $q$ is a model selection problem and a poor model choice may result in an incomplete representation of the data. A suitable choice is made through domain knowledge or using data exploration together with some available heuristic options \citep{valle1999selection,saporta2009principal}.




\subsection{Intrusion detection system based on PCA}
\label{subsection2.1}
As mentioned above, an unsupervised NIDS operates in two phases: the training phase and the monitoring phase. In the training phase, the normal activities patterns are learned through fitting a model to the recorded clean dataset (referred to as training set of data). Precisely, assume that $p$ features are measured for each of the $n$ normal activities connections. Let $y_{ij}$ be the recorded observation of the $i$th normal activity connection made on the $j$th feature contained in a $n\times p$ training data matrix ${Y}$. The columns of the matrix $Y$ are standardized to have zero mean and unit standard deviation, and the standardized training dataset is stored in a matrix $X$, that is, the elements of $X$ are obtained by $x_{ij}=(y_{ij}-\bar{y}_j)/{s_j}$, where $\bar{y}_j$ and $s_j$ are the mean and standard deviation of the $j$th column of matrix $Y$, respectively. A PCA model given in \eqref{svd} is fitted to the data in $X$.

In the monitoring phase, a newly observed set of network traffic connections (referred to as test set of data) is checked with the normal activities model learned in the training phase and an anomaly warning is triggered if any of the new connections shows deviation from the trained model. 

More formally, let $Y^f$ be an $m\times p$ matrix of newly observed set of connections. To obtain $X^f$, we adjust each element $y^f_{ij}$ of the matrix $Y^f$ by first subtracting the mean of the $j$th column of the normal activities dataset $Y$ and then divide by the standard deviation of the $j$th column of normal activities dataset $Y$ that is  $x^f_{ij}=(y^f_{ij}-\bar{y}_j)/s_j$. \cite{shyu2003novel} develop an intrusion detection system based on the statistic $t_i$ in \eqref{ti}. They propose to monitor two statistics: one based on principal components (first few components that account for most of the variation in the data) and another based on minor components (last few components). To obtain a threshold for monitoring future data, they rely on the empirical distribution of $t_i$ (they have not mentioned any specific estimation method) since the assumption of multivariate normality is often violated in network connections data. Since different kinds of attacks might create shifts (anomalies) along different eigenvectors, the disadvantage of this approach is that it might miss signals along the eigenvectors that are disregarded. \cite{wang2006identifying} proposed to monitor the error induced by the approximate reconstruction of the actual connection using the expression in \eqref{appx} instead of monitoring two statistics; that is, they monitor
$$
\varepsilon=\vert\vert X^f-\tilde{X}^f\vert\vert^2,
$$
where $\tilde{X}^f=X^f\tilde{V}\tilde{V}^T$ and $\tilde{V}$ is based on the normal activities dataset obtained in \eqref{svd}. We will refer to the method of \cite{shyu2003novel} as PCC and that of \cite{wang2006identifying} as WBPCA for the rest of this paper. \cite{wang2006identifying} have shown numerically that WBPCA has a slightly better performance than PCC in terms of detection rate and false discovery rate. The downside of WBPCA is that it only takes into account the principal components and ignores the minor components completely; therefore, their monitoring models are likely to be less sensitive to the attacks that affect only the minor components and anomalies that are rare events.

It is possible that a particular type of attack creates a shift along the directions of fewer eigenvectors (possibly a mix of some principal and some minor components) and there is negligible or no change along the other dimensions. This is because each PC is a linear combination of the $p$ observed variables and it is common in practice that a particular PC has larger coefficients for some of the $p$ observed variables while the coefficients for the remainder of the observed variable are close to zero. Now if an anomalous connection differs to the normal activity connections with respect to only one feature (say F) and only one particular PC (say PCF) has a large coefficient for F (the coefficient for the remainder of the observed variables may or may not be large), the shift caused by this anomalous connection will be expressed only along PCF. In this case, if $t_i$ is based on only the PCF, the anomaly will be more likely to be detected, compared to if $t_i$ is based on the rest of the components. Further, the first few PCs that account for most of the variation in the data are noisier and the possibility is that the perturbations caused by anomalies might be less pronounced along these PCs, compared to the PCs that account for less variation in the data. In order to make the system more sensitive to such anomalies, it is important to take into account the most affected dimensions rather than choosing some of the principal or minor components which (if not affected) will add noise and may actually obscure the anomalies. 

We propose a similar system to PCC that is based on PCA but unlike the above mentioned approaches we take into account the dimensions that may be most affected by anomalies or attacks. To proceed, re-write the expression in \eqref{svd} as
\begin{equation}
XV\Gamma^{-1}=U.
\label{svd1}
\end{equation}
Since the columns of $U$ are orthonormal, we have
\begin{equation}
(n-1)\textrm{cov}(XV\Gamma^{-1})=(n-1)\textrm{cov}(U)=I,
\label{cov1}
\end{equation}
where $I$ is an identity matrix.
If $X$ and $X^f$ are generated by the same model then the same relationship in \eqref{cov1} holds for $X^f$, that is,
\begin{equation*}
(n-1)\textrm{cov}(X^fV\Gamma^{-1})=(n-1)\textrm{cov}(U^f)=I.
\end{equation*}
On the other hand, if $X^f$ is generated by a different model then all or some of the diagonal elements of $(n-1)\textrm{cov}(X^fV\Gamma^{-1})$ will be larger than unity. It is possible that some of the diagonal elements of $(n-1)\textrm{cov}(X^fV\Gamma^{-1})$ are less affected or not affected at all. In this situation, it is more sensible to evaluate the standard deviations of the columns of $\sqrt{n-1}U^f$, say $s_j^u$ (the diagonal elements of $(n-1)\textrm{cov}(X^fV\Gamma^{-1})$), and consider those dimensions for which the corresponding $s_j^u$'s deviate from unity.

In our monitoring system, we exploit the above fact and consider only those components that are affected by the anomalies, thereby making the system more sensitive to the deviations from the normal activities model. Let $r_j^u$ be the standard deviation of the $j$th column of $\sqrt{n-1}U$. To account for sampling variation in each $s_j^u$, we use a threshold $\delta_{j}^{(1-\alpha)}$ obtained as the $(1-\alpha)$th quantile of the bootstrap distribution of $r_j^u$. The $j$th component is considered affected if $s_j^u$ exceeds $\delta_{j}^{(1-\alpha)}$ and is thus considered towards the final score  $t_i^f$ of the newly observed data. To identify the anomalous connection, we use another threshold $\theta^{(1-\alpha)}$ obtained as the $(1-\alpha)$th quantile of $\chi^2_q$, where $q$ is the number of $s_j^u$ that exceed $\delta_{j}^{(1-\alpha)}$.
Our system is then based on the anomaly affected components.

We call this proposal AAD and formalize it in the following algorithm:
\begin{enumerate}
\item Obtain $X$ by standardizing the normal activities dataset $Y$ that is $x_{ij}=(y_{ij}-\bar{y}_j)/{s_j}$.
\item Estimate $V$ and $\Lambda$ based on $X$ using singular value decomposition.
\item Obtain $\delta_{j}^{(1-\alpha)}$ based on the bootstrap distribution of $r_j^u$.
\item Obtain $U^f$ using the newly observed dataset that is $U^f=X^fV\Gamma^{-1}$, where $x^f_{ij}=(y^f_{ij}-\bar{y}_j)/s_j$.
\item Calculate $s_j^u$'s, the standard deviations of the columns of $\sqrt{n-1}U^f$. In the absence of intrusion, we expect all $s_j^u$'s to be unity. If $s_j^u > \delta_{j}^{(1-\alpha)}$ for any $j$, then declare the possible presence of anomalous connections in $Y^f$.
\item To identify the anomalous connections, calculate $t_i^f$ using the expression in \eqref{ti} that is $t_i^f = (n-1)\sum_{j=1}^q (u_{ij}^f)^2$, where the sum is performed over the $q$ affected columns of $U^f$.
\item For the $i$th connection, if $t_i^f > \theta^{(1-\alpha)}$, where $\theta^{(1-\alpha)}$ is the $(1-\alpha)$th quantile of $\chi^2_q$, then declare it anomalous.
\end{enumerate}
It is important to reiterate here that AAD relies on the assumption of multivariate normality for its threshold $\theta_{\alpha}$. As noted above, this assumption is often violated in network data and may render AAD unreliable. One solution is to consider the bootstrap distribution of $t_i$ which requires upfront knowledge of the components whose $s_j^u$ exceeds $\delta^{(1-\alpha)}$. In our use of AAD, we rely on the bootstrap distribution of $\theta^{(1-\alpha)}$ unless otherwise stated. Making available a computationally cumbersome bootstrap distribution of $t_i$ in real time --- after knowing the important components that need to be evaluated --- may impede the real time detection in large-scale networks. In such situations, one could use the empirical distribution of $t_i$ as this is computationally faster to evaluate.

Another possibility is to use
\begin{equation}
t_i^{fw} = (n-1)\sum_{j=1}^q w_j u_{ij}^2,
\label{tiw}
\end{equation}
where $w_j=s_j^u$. In the absence of any intrusion or anomaly, we have $s_j^u=r_j^u=1$; therefore, both $t_i$ and $t_i^f$ follow the same distribution. However, some of the $s_j^u$'s will be larger than unity in the presence of unusual connections in $X^f$. Giving more importance to the components that are affected by weighting them with corresponding $s_j^u$'s will make the anomalies more pronounced and will result in a shift in $t_i^{fw}$ relative to $t_i$ for all anomalous connections in $X^f$. In the rest of this paper, we refer to a detection system based on \eqref{tiw} as WAAD and describe it via the following algorithm:
\begin{enumerate}
\item Obtain $X$ by standardizing the normal activities dataset $Y$ that is $x_{ij}=(y_{ij}-\bar{y}_j)/{s_j}$.
\item Estimate $V$ and $\Lambda$ based on $X$ using singular value decomposition.
\item Obtain $U^f$ using the future activities dataset that is $U^f=X^fV\Gamma^{-1}$, where $x^f_{ij}=(y^f_{ij}-\bar{y}_j)/s_j$.
\item Calculate $s_j^u$'s, the standard deviations of the columns of $\sqrt{n-1}U^f$. In the absence of intrusion, we expect all $s_j^u$'s to be unity. If $s_j^u > \delta_{j}^{(1-\alpha)}$ for any $j$, then declare the possible presence of anomalous connections in $Y^f$.
\item To identify the anomalous connections, calculate $t_i^{fw}$ using the expression in \eqref{tiw} that is $t_i^{fw} = (n-1)\sum_{j=1}^p w_j (u_{ij}^f)^2$.
\item For the $i$th connection, if $t_i^{fw} > \theta^{(1-\alpha)}$, where $\theta^{(1-\alpha)}$ is the $(1-\alpha)$th quantile of the bootstrap distribution of $t_i^{w} = (n-1)\sum_{j=1}^q w_j (u_{ij})^2$, then declare it anomalous.
\end{enumerate}
Like AAD, WAAD requires the bootstrap distribution of $t_i^{w}$, which in turn requires the upfront knowledge of the $s_j^u$. The WAAD also consider the unaffected components (although with lower weights) in the calculation of $t_i^w$ and $t_i^{fw}$, which makes it noisier; therefore, it may not be as effective in identifying anomalies as the AAD.

\section{Simulation study}
\label{sunsection3}
The goal of this section is to numerically evaluate the performance of our proposed methods for detecting anomalies. We also demonstrated the effectiveness of the proposed methods in comparison to the WBPCA while considering a number of factors that might be affecting the performance of the methods. In the presence of true mean $\mu$ and covariance $\Sigma$, we included a Mahalanobis distance approach \citep{koeman2019critical} based on $\mu$ and $\Sigma$ as a gold standard. The squared Mahalanobis distance between a sample $x_i^f$ and $\mu$ given a true covariance matrix $\Sigma$ is given by
$$
D_i^2=(x_i^f-\mu) \Sigma^{-1} (x_i^f-\mu)^T.
$$
In the absence of anomalous observations and under the assumption of normality and independence of observations, the quantity $D^2$ follows an exact chi-square distribution with $p$ degrees of freedom; that is, $D_i^2\sim \chi^2_{(p)}$ \citep{johnson2002applied}. In the rest of the paper we will refer to this approach as TRUE.

We generated a clean dataset $Y$ of size $n$ from a $p$-dimensional multivariate normal distribution with mean vector $\mu$ and a covariance matrix $\Sigma$. For the clean dataset, without loss of generality, we set $\mu=0$ and considered $\Sigma=\Sigma_\rho$ having a first-order autoregressive correlation structure, AR(1); that is, the elements of $\Sigma_\rho$ are given by
$$
\sigma_{ij}=\rho^{|i-j|}~~~~\textrm{for}~~ 1\leq i,j \leq p,
$$
where $\rho\in [0,1)$ is a parameter that determines $\Sigma$. A smaller value of $\rho$ shifts variance towards minor components. For example, $\Sigma_0$ equals an identity matrix in which case the variance is uniformly distributed over $p$ components. On the other hand, a larger value of $\rho$ shifts more variance towards the principal components in which case the first few eigenvalues will explain most of the variation.

To generate a set of anomalous observations, we followed the simulation setup provided in \cite{warton2008penalized}, which allows a shift along a set of chosen eigenvectors. We chose a smaller set of observations from $n$ normal observations and made them anomalous by adding a constant of size $\eta=c \sqrt{\lambda_j}$, where $c\in\{1,2,3\}$, along a few chosen eigenvectors from $V$. This set of anomalous observations was then used to contaminate another set of $m$ normal observations $Y^f$ randomly chosen from training dataset. For example, an observation $y_{i}^f$ was anomalous if it was reconstructed after a shift is created along some of the $v_{j}$'s (the columns of $V$) by adding a constant $\eta$ to the corresponding $z_{ij}$ (the $ij$th element of $Z$).

We used the AAD and WAAD as detailed in Section \ref{subsection2.1} in comparison to WBPCA and TRUE to detect and identify the $m$ anomalous observations. For WBPCA, we calculated the eigenvalues of correlation matrix and retained components using ``eigenvalue greater than one rule''. We used the Receiver Operating Characteristic (ROC) curve, a plot of the true positive rate against the false positive rate, as a metric to evaluate the performance of the methods \citep{ringberg2007sensitivity}.

\begin{figure}[!t]
\begin{center}$
\begin{array}{c}
\includegraphics[scale=0.15,angle=270]{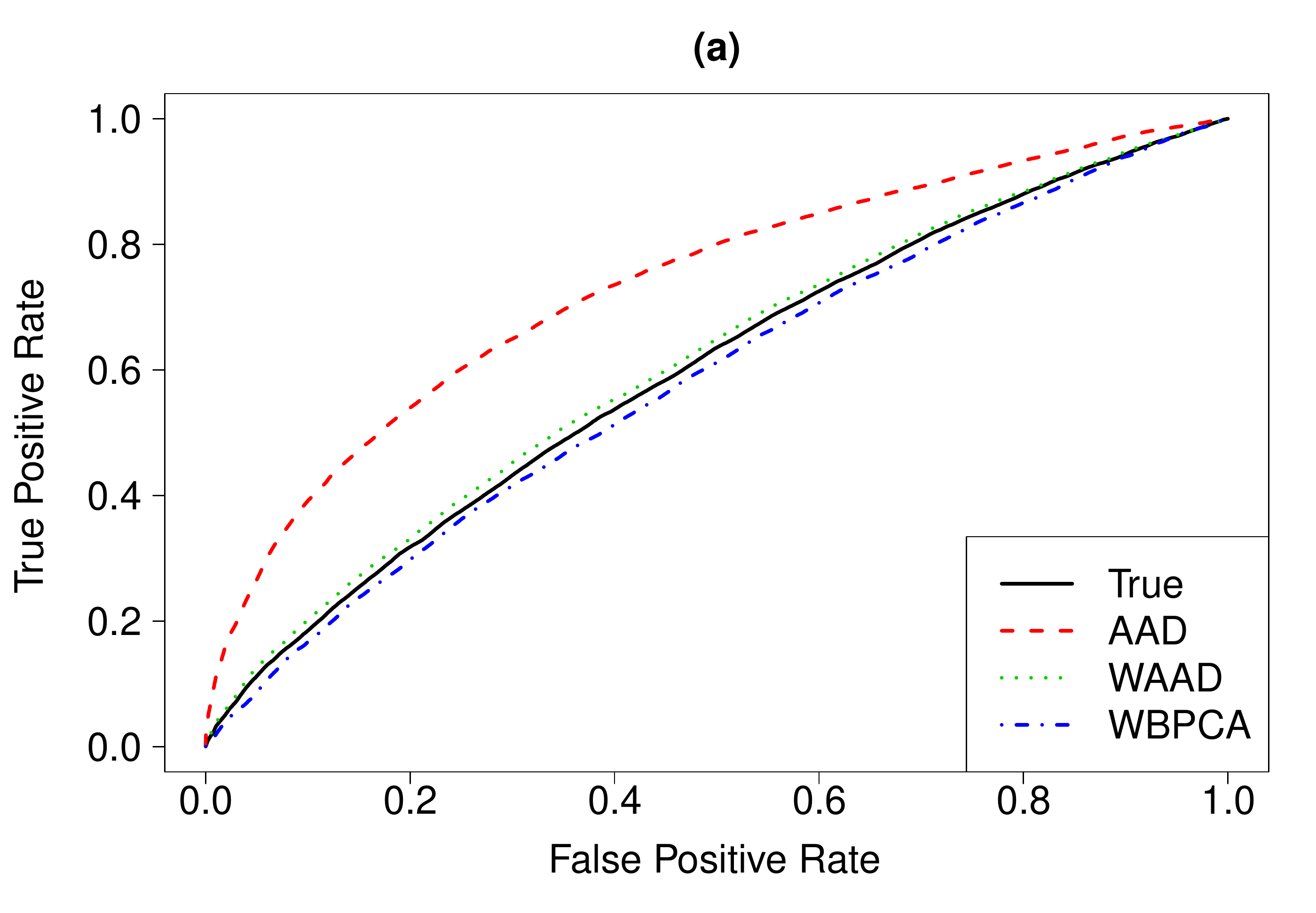}
\includegraphics[scale=0.15,angle=270]{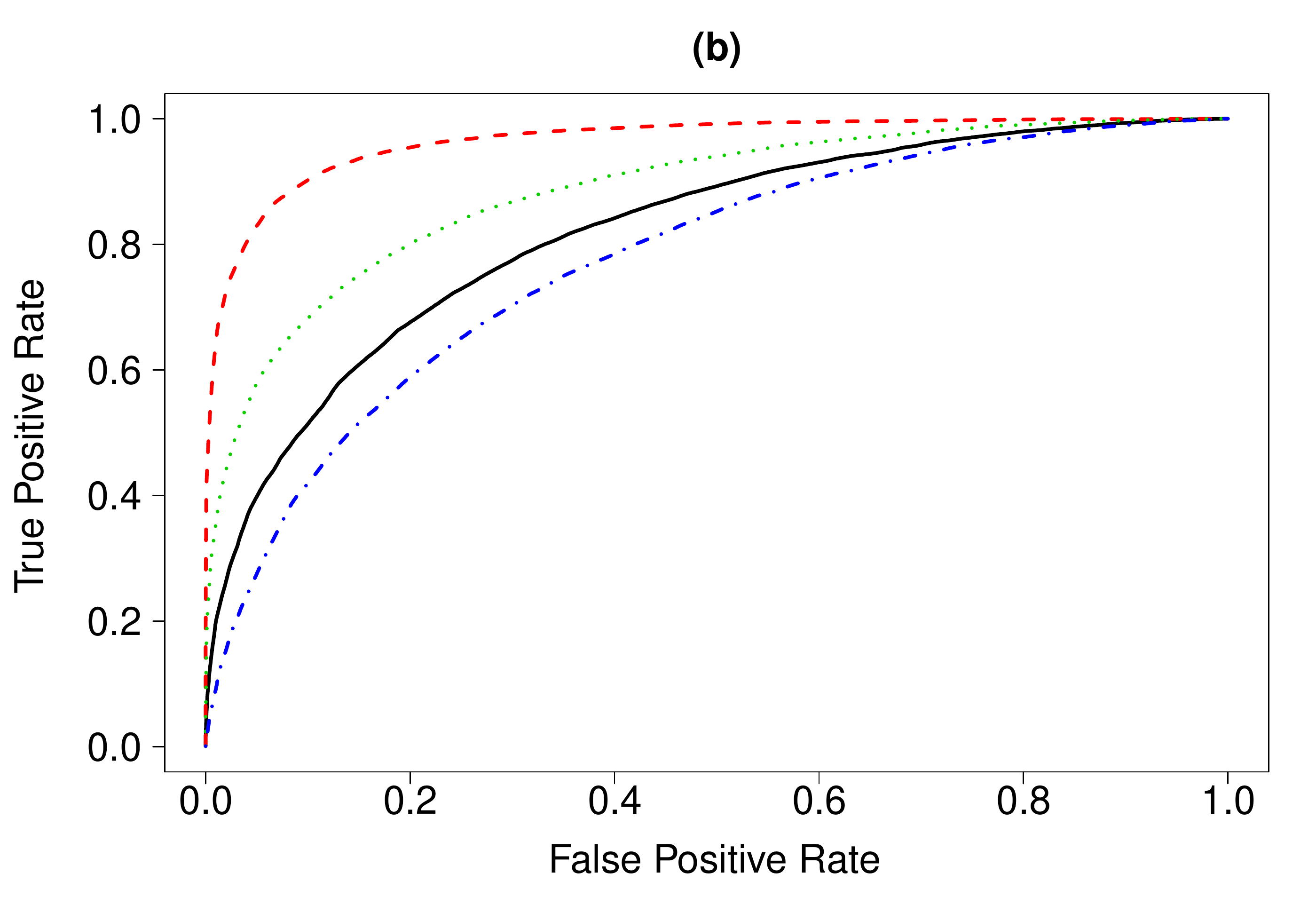}\\
\includegraphics[scale=0.15,angle=270]{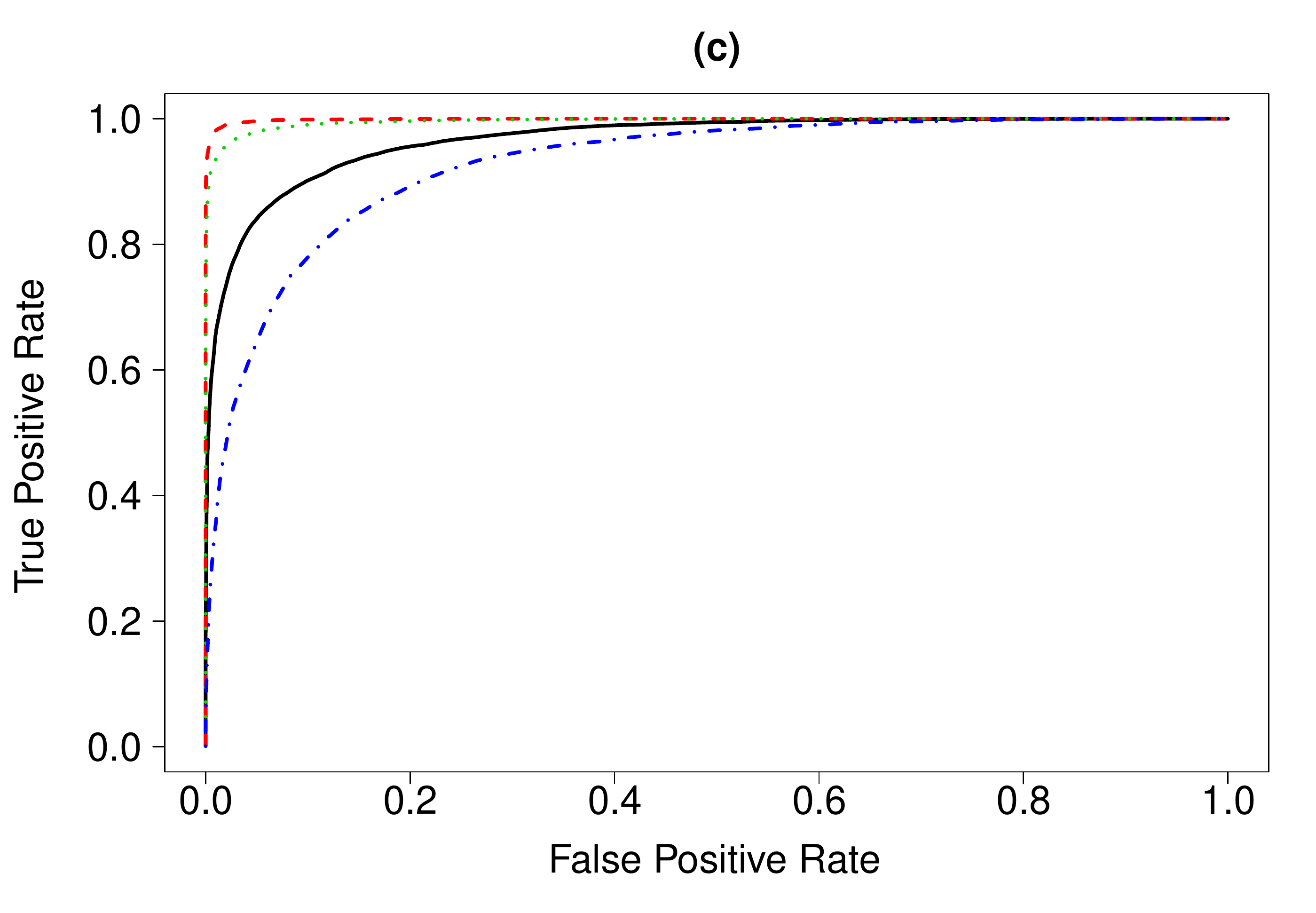}
\end{array}$
  \caption{ROC curves. A sample of size $n=10,000$ is drawn from a $30$-dimensional normal distribution with $\mu=0$ and $\Sigma_{0.9}$ to train a PCA model. A set of 100 anomalous observations are created to contaminate another set of $m=5000$ normal observations by adding a constant of size $\eta=c\sqrt{\lambda_j}$: (a) $c=1$, (b) $c=2$ and (c) $c=3$, to shift them along a random set of three distinct eigenvectors. The results are averaged over 1000 independent experiments. In each experiment the set of three eigenvectors along which the shifts are created is allowed to vary.}\label{simroc1}
\end{center}
\end{figure}

\begin{figure}[!t]
\begin{center}$
\begin{array}{c}
\includegraphics[scale=0.15,angle=270]{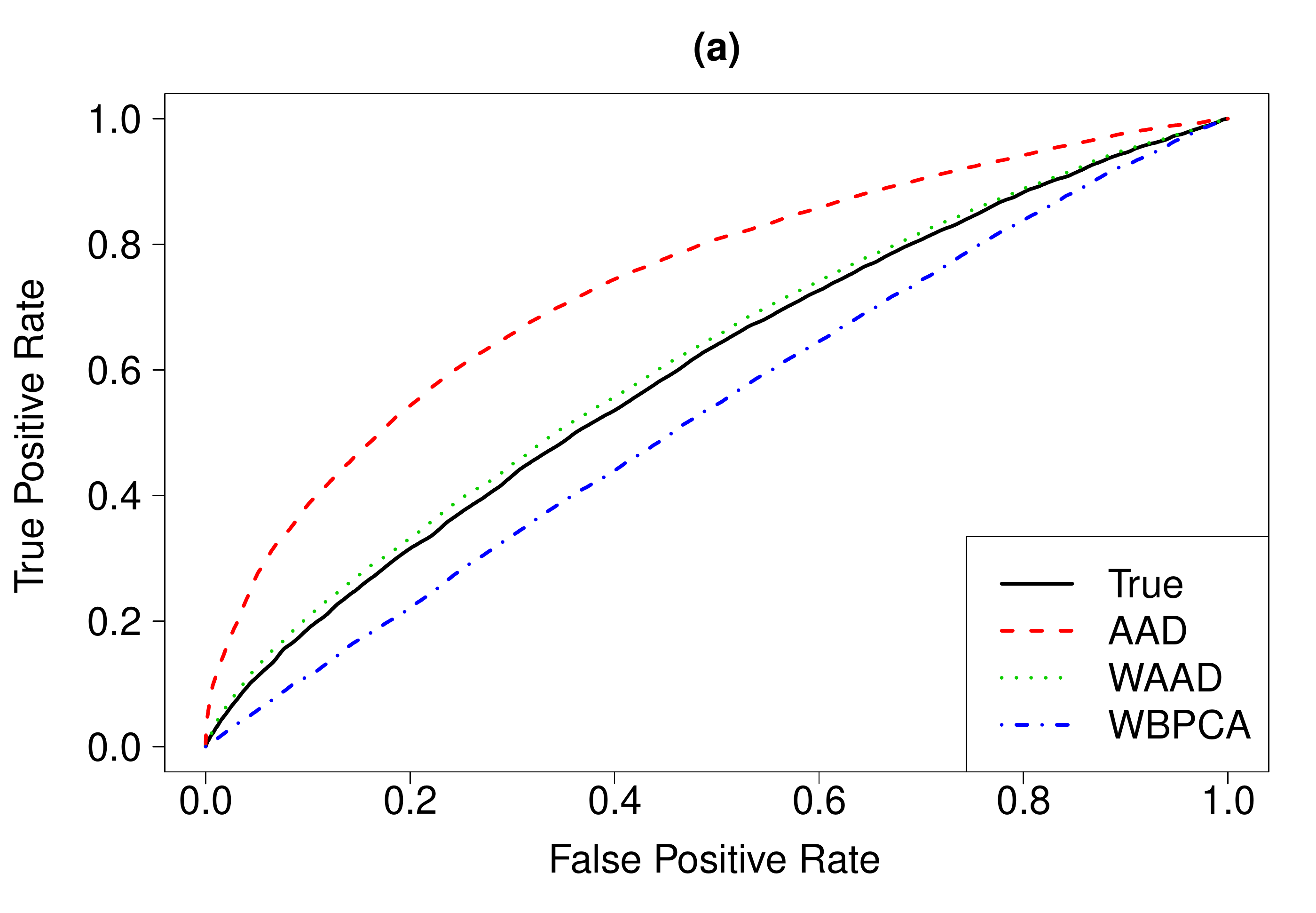}
\includegraphics[scale=0.15,angle=270]{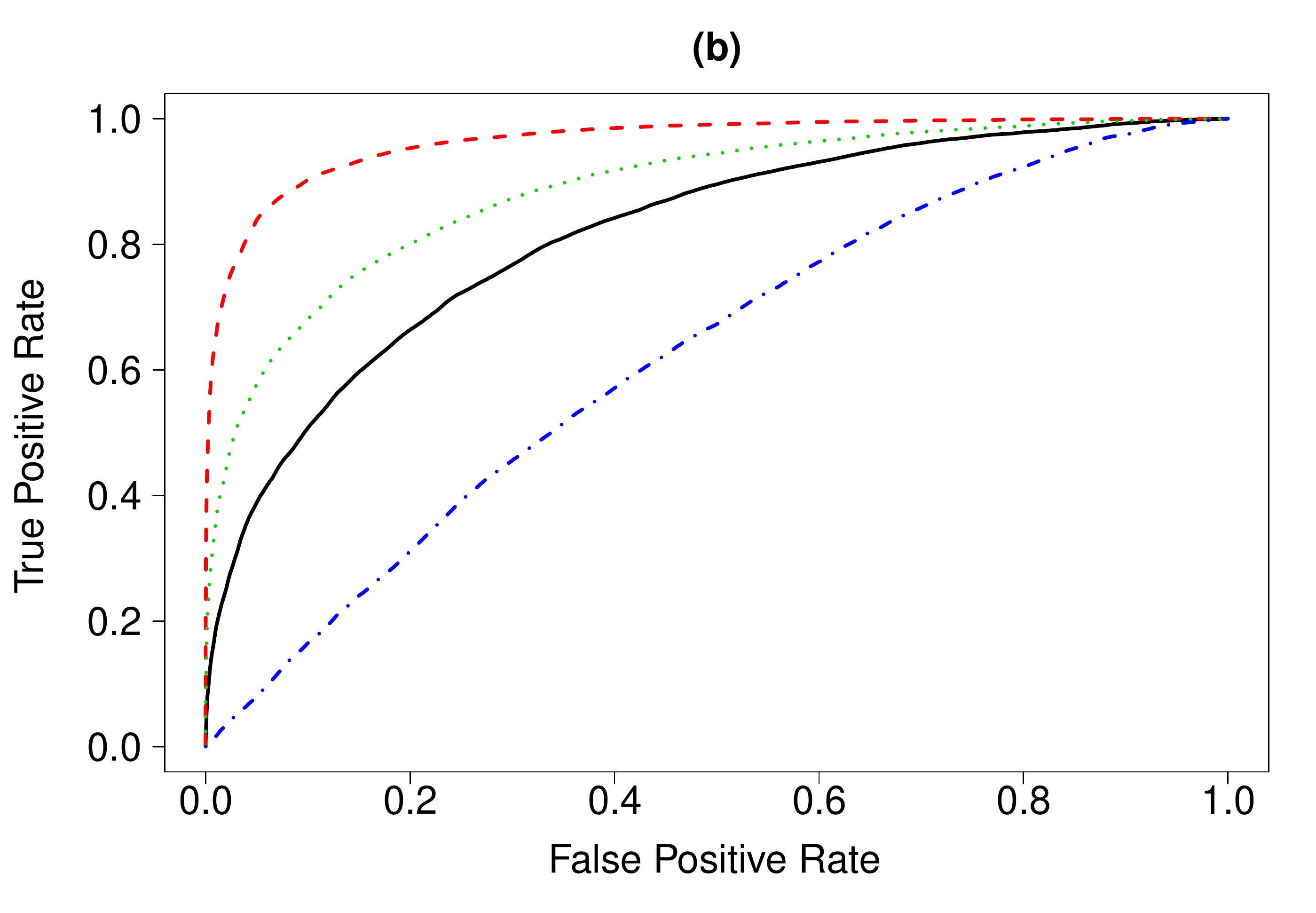}\\
\includegraphics[scale=0.15,angle=270]{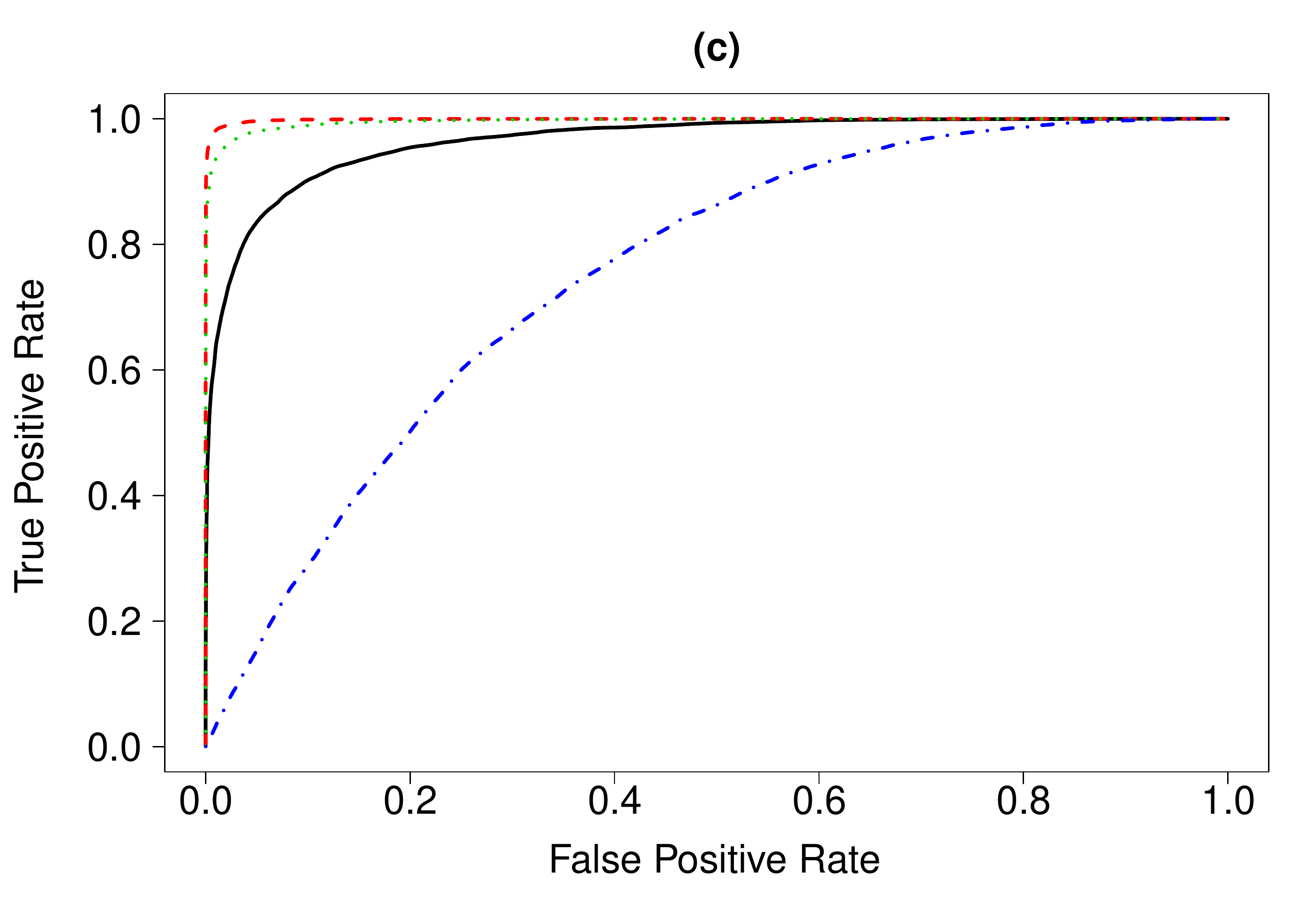}
\end{array}$
\caption{ROC curves. A sample of size $n=10,000$ is drawn from a $30$-dimensional normal distribution with $\mu=0$ and $\Sigma_{0.9}$ to train PCA model. A set of 100 anomalous observations are created to contaminate another set of $m=5000$ normal observations by adding a constant of size $\eta=c\sqrt{\lambda_j}$: (a) $c=1$, (b) $c=2$ and (c) $c=3$, to shift them along the last three eigenvectors. The results are averaged over 1000 independent experiments.}\label{simroc2}
\end{center}
\end{figure}

\begin{figure}[!t]
\begin{center}$
\begin{array}{c}
\includegraphics[scale=0.15,angle=270]{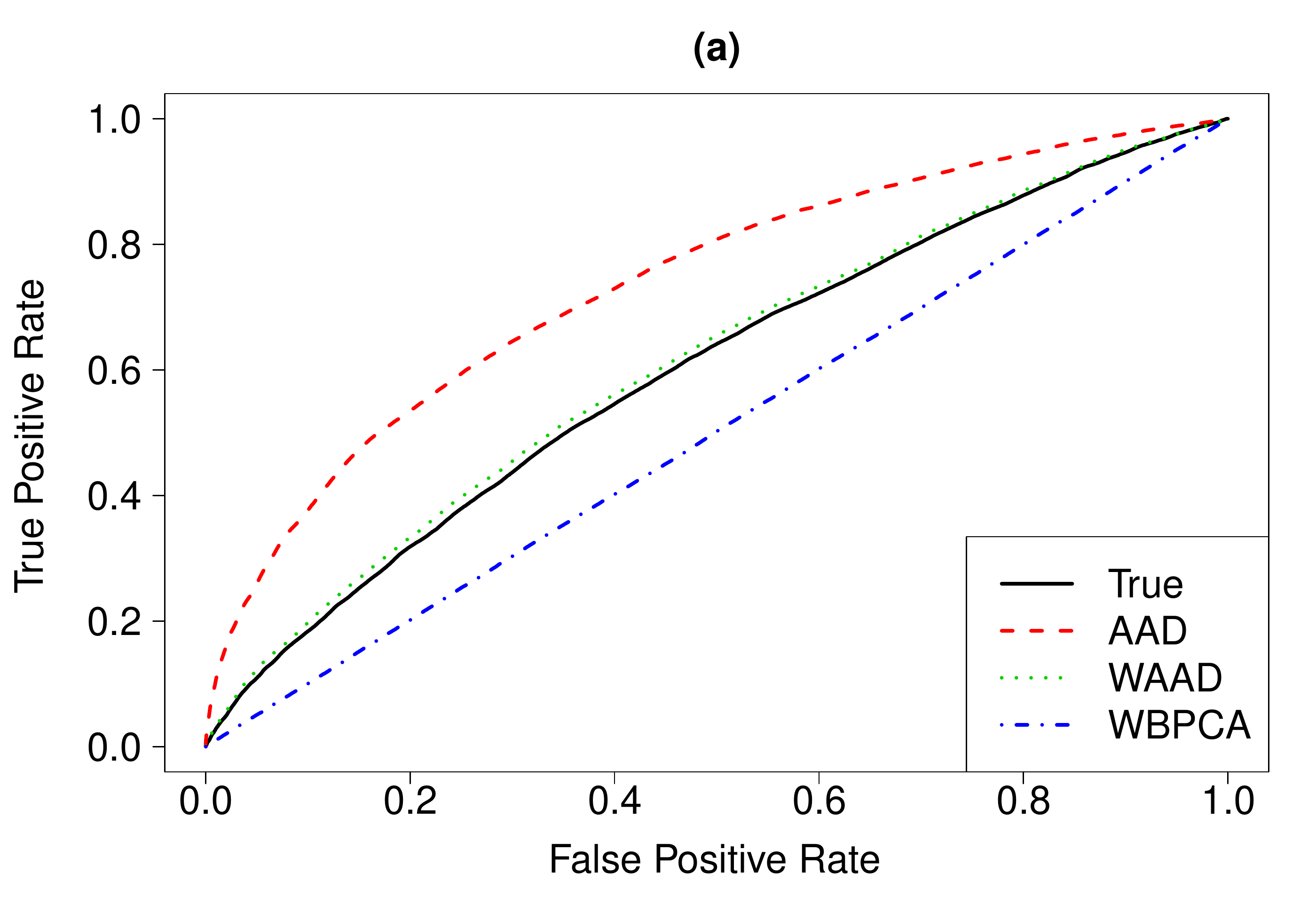}
\includegraphics[scale=0.15,angle=270]{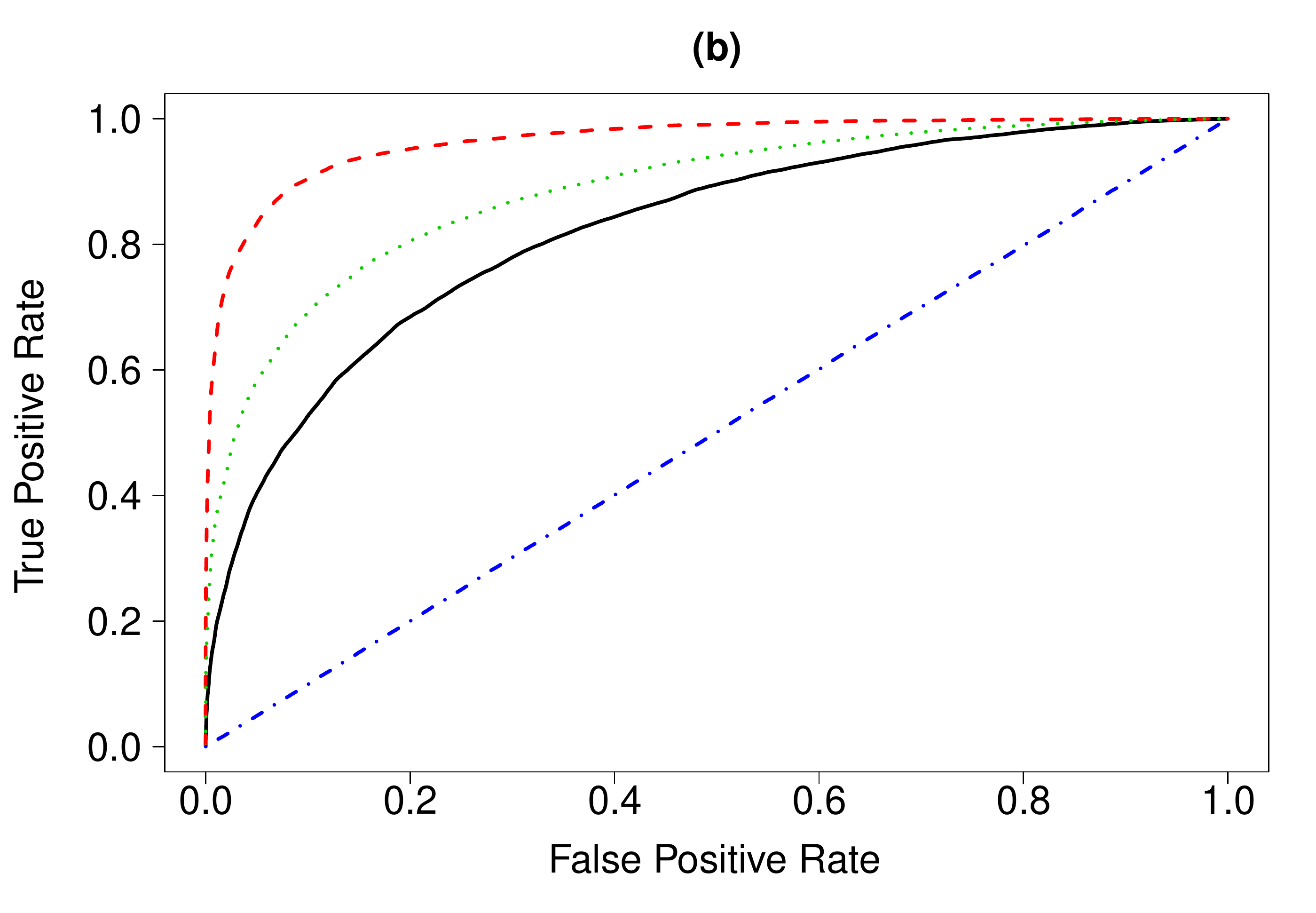}\\
\includegraphics[scale=0.15,angle=270]{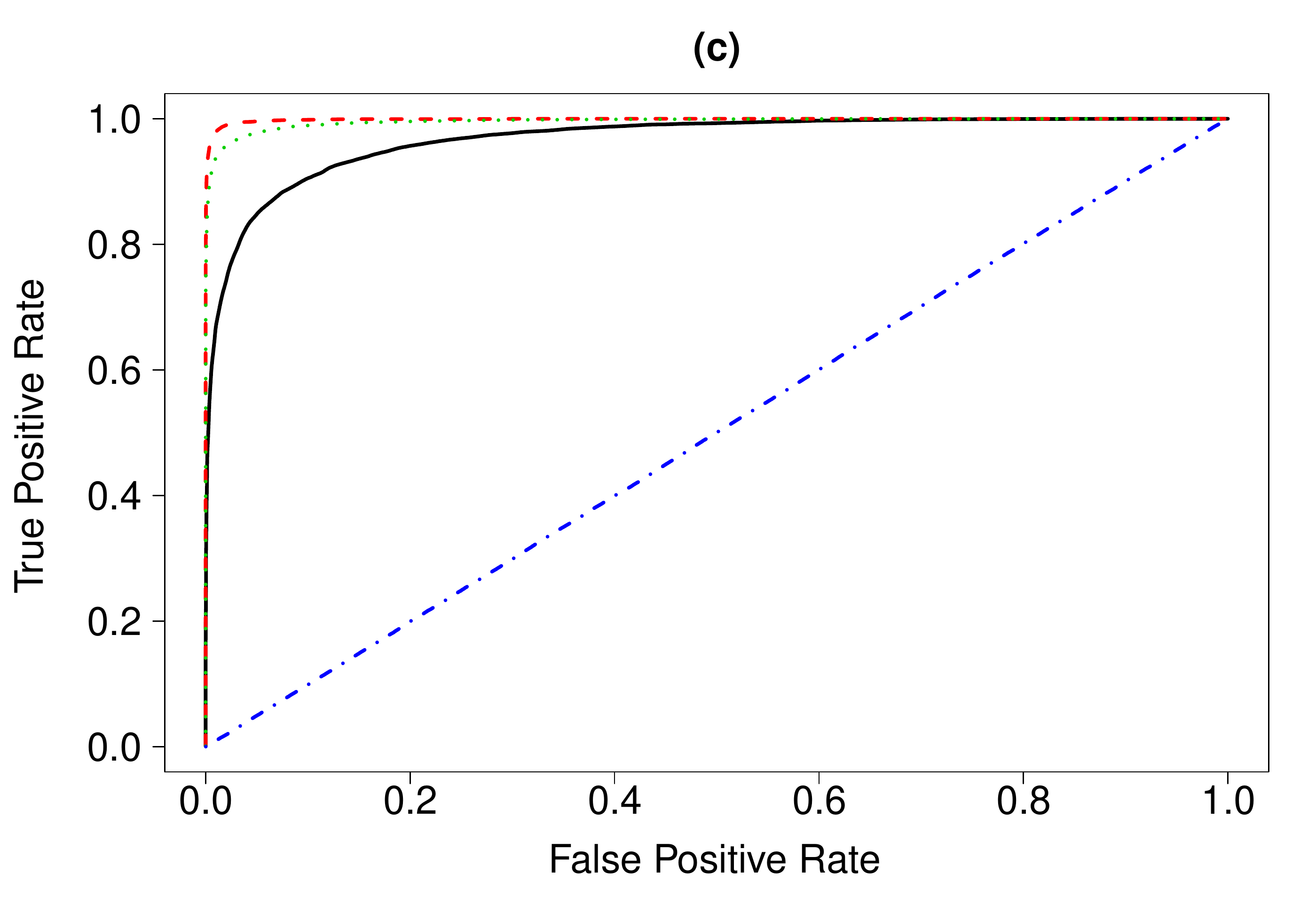}
\end{array}$
\caption{ROC curves. A sample of size $n=10,000$ is drawn from a $30$-dimensional normal distribution with $\mu=0$ and $\Sigma_{0.9}$ to train PCA model.A set of 100 anomalous observations are created to contaminate another set of $m=5000$ normal observations by adding a constant of size $\eta=c\sqrt{\lambda_j}$: (a) $c=1$, (b) $c=2$ and (c) $c=3$, to shift them along the first three eigenvectors. The results are averaged over 1000 independent experiments.}\label{simroc3}
\end{center}
\end{figure}

For the experiments, the results of which are shown here, we set $p=30$, $\rho=0.9$, and created anomalous observations by choosing 100 normal observations at random and shifted them along three chosen eigenvectors. We then contaminated $m=4900+100=5000$ observations chosen randomly from $n=10,000$ normal observations with the 100 anomalous observations. The average ROC curves for 1000 independent experiments are presented in Figures \ref{simroc1}, \ref{simroc2} and \ref{simroc3}. The three eigenvectors along which the shifts were created were randomly selected in each of the 1000 independent experiments (see results in Figure \ref{simroc1}), were fixed to the last three eigenvectors (see results in Figure \ref{simroc2}) and were fixed to the first three eigenvectors (see results in Figure \ref{simroc3}). 

Overall, the detection rate increased with increasing size of the shift and the results clearly indicated the superior performance of AAD in terms of achieving higher detection rate with much lower false alarm rate. Further, AAD performed consistently well regardless of the dimension of the created shift. Its performance was even better than the TRUE. The reason is that the Mahalanobis distance also aggregates across all variables including irrelevant variables, hence masking the outlying effect. Similar to AAD, the performance of WAAD was also consistent across the three simulation scenarios; however, it was still polluted by the aggregated noise in $t_i^{fw}$ due to the unaffected dimensions.

The effectiveness of WBPCA varied depending on the chosen dimension along which the shift was created.
The WBPCA detection rate was lower if the shift was along the principal components (see Figure \ref{simroc3}). However, its performance improved when the shift was created along the minor components (see Figure \ref{simroc2}). This is because most of the error variation lies along the dominant eigenvectors when $\rho$ is close to one which makes it difficult to detect anomalous observations.


\section{Real data analysis}
\label{section4}
In this section, we demonstrate the performance of our proposed methods and those from the literature through the analysis of two real network datasets, namely the UNSW-NB15 dataset and KDD'99 dataset.
\subsection{UNSW-NB15 dataset}
\label{subsection4.1}

A real UNSW-NB15 dataset was recently released by the Australian Centre for Cyber Security.
The dataset is described in \cite{moustafa2015unsw} and analyzed (using a variety of methods) in \cite{moustafa2016evaluation} and \cite{moustafa2017novel}. It is publicly available for research purposes\footnote{\url{https://www.unsw.adfa.edu.au/unsw-canberra-cyber/cybersecurity/ADFA-NB15-Datasets/}}.
This dataset has 49 features and a total of 2,540,044 connection records including a variety of attacks. The 49 features also include 12 features that are derived from the rest of 35 directly measured features and the two attributes that represent class labels.
We do not use the 12 derived features and also exclude seven nominal features (Source IP address, Destination IP address, Source port number, Destination port number, transaction protocol, service and state).
All the records are labelled as normal records or attacks using the IXIA PerfectStorm tool\footnote{\url{https://www.ixiacom.com/products/perfectstorm}} that contains the signatures of naive attacks and is continuously updated from a CVE site\footnote{\url{https://cve.mitre.org/}}.
The attacks are further categorized into their types: Fuzzers, Analysis, Backdoors, DoS, Exploits, Generic, Reconnaissance, Shellcode, Worms \citep[see][for the description of each type of attack]{moustafa2015unsw}. Some of the above mentioned nominal variables could be included in the analysis in the form of binary indicator variables since they have a small number of categories in the available data (transaction protocol has 135 categories, service has 16 categories and state 13 categories). However, the number of categories may vary in practical applications.

As mentioned earlier, the key requirement of an unsupervised NIDS is the availability of a normal activities dataset. In the UNSW-NB15 dataset, the abnormal traffic is introduced synthetically at particular time points, so we have a large portion of the data that is not affected by the abnormal traffic (i.e. observations 186,789 to 1,087,248, which we will refer to as clean data). We used the observations from 300,001 to 900,000 of the clean data to train the PCA model and to estimate $\delta_{j}^{(1-\alpha)}$. The data observed before 300,001 and after 900,000 are used as a test set of data.

\begin{figure}[!t]
\centering
\subfloat{\includegraphics[scale=0.300]{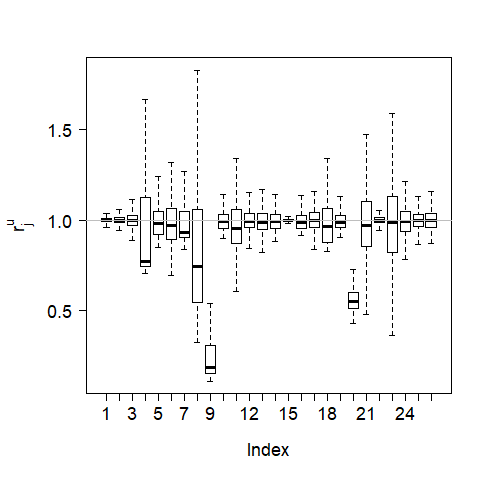}}\\
\caption{Plot of the values of $r_j^u$'s obtained using 5000 bootstrap samples of size 10000 from the training set of the UNSW-NB15 dataset. The horizontal gray line indicates the expected value of $r_j^u$ in the absence of anomalous activities. }\label{UNSW-NB15-maxdeltaout}
\end{figure}

\begin{figure}[!htb]
\centering
\subfloat{\includegraphics[scale=0.300]{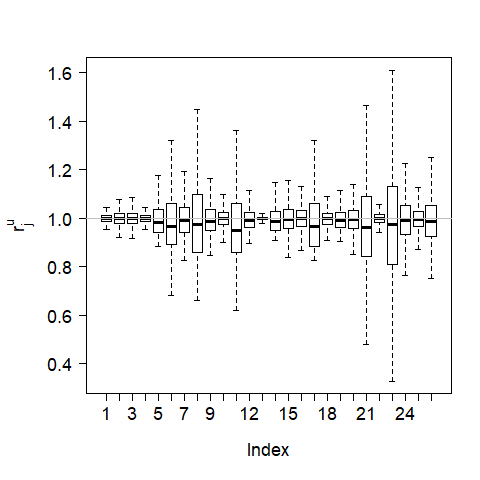}}\\
\caption{Plot of the values of $r_j^u$'s obtained using 5000 bootstrap samples of size 10000 from the training set of the UNSW-NB15 dataset after removal of 79 unusual connections. The horizontal gray line indicates the expected value of $r_j^u$ in the absence of anomalous activities. }\label{UNSW-NB15-maxdelta}
\end{figure}

We used 5,000 bootstrap samples of size 10,000 to construct the distributions of $r_j^u$'s (see Figure \ref{UNSW-NB15-maxdeltaout}). The bootstrap distributions of $r_j^u$'s were expected to be centered at one. At this point it is important to notice that the network connections data are usually skewed which may make the bootstrap distributions of $r_j^u$'s slightly off from the expected center of one. However, misplacement of the entire distribution of $r_j^u$ is an indication of the presence of outliers in the data. The presence of large outliers in the training data has the potential to make a NIDS less sensitive. We noticed peculiarities in the distributions of some of the $r_j^u$'s, particularly the distributions of $r_8^u$, $r_9^u$ and $r_{20}^u$ were far-off from the expected center. This indicated the presence of some unusually large observations of the variables in the training data that were loading heavily on PC-8, PC-9 and PC-20. The PC-8 had loadings of 0.91 and 0.19, respectively, for `source bits per second' and `destination bits per second', PC-9 had loadings of 0.21 and 0.97, respectively for `source bits per second' and `record total duration' (loading for all other variable were less than 0.1) and the PC-20 had a large loading of -0.70 and 0.71, respectively, for `source jitter' and `destination jitter'. By further inspection we found six unusually large values of the `record total duration', 39 unusually large values of the `source bits per second', 3 unusually large values of the `destination bits per second', and 31 unusually large values of the `source jitter'. Most of the 31 connections that had unusually large values for the `source jitter' had also unusual values for the `destination jitter'. The removal of these 79 unusual observations considerably alleviated the peculiarities in the distributions of $r_j^u$ (see Figure \ref{UNSW-NB15-maxdelta}). 



\begin{figure}[!htb]
\centering
\subfloat{\includegraphics[scale=0.300]{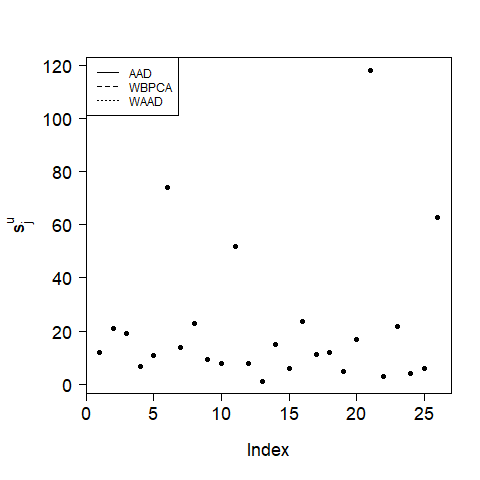}}
\caption{Plot of $s_j^u$'s for UNSW-NB15 dataset. The whole UNSW-NB15 dataset was examined at once.}\label{UNSW-NB15-delta}
\end{figure}

\begin{figure}
\begin{center}$
\begin{array}{c}
\includegraphics[scale=0.250]{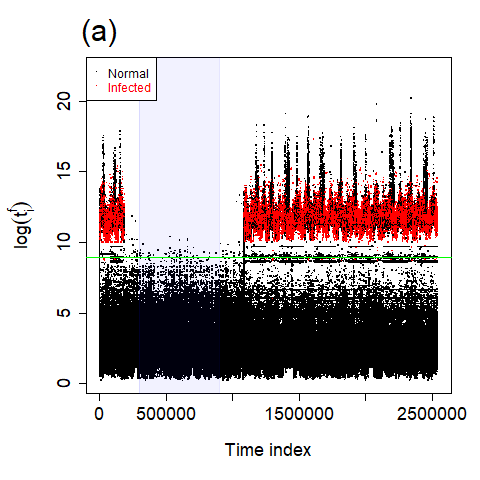}
\includegraphics[scale=0.250]{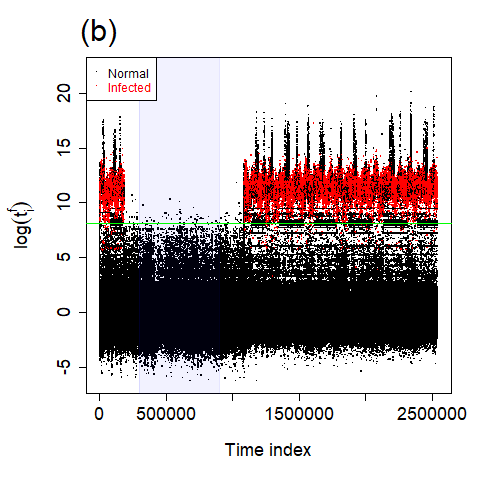}\\
\includegraphics[scale=0.250]{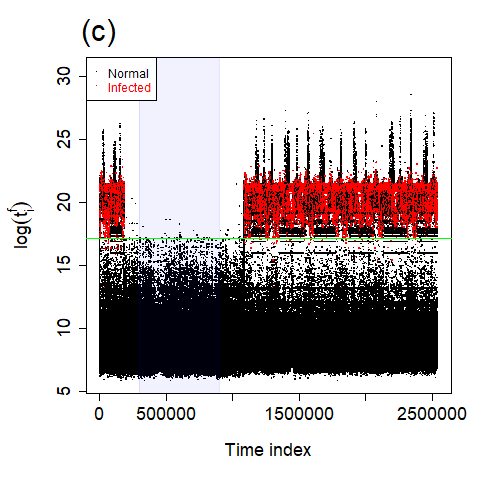}
\includegraphics[scale=0.250]{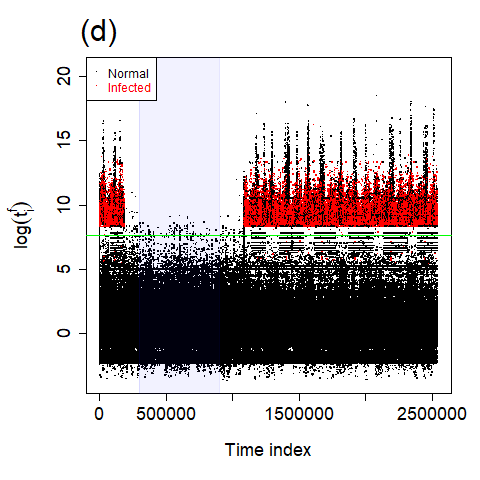}
\end{array}$
\end{center}
\caption{Plots of the score statistic $t_i^f$ for the UNSW-NB15 dataset: (a) AAD, (b) AAD based on the four most affected dimensions (see Figure \ref{UNSW-NB15-delta}), (c) WAAD and (d) WBPCA. The horizontal green line indicates the threshold value $\theta_{\alpha=0.0001}$. The training dataset is shaded in light-blue.}\label{UNSW-NB15-f1}
\end{figure}
\begin{figure}
\centering
\subfloat{\includegraphics[width=5cm]{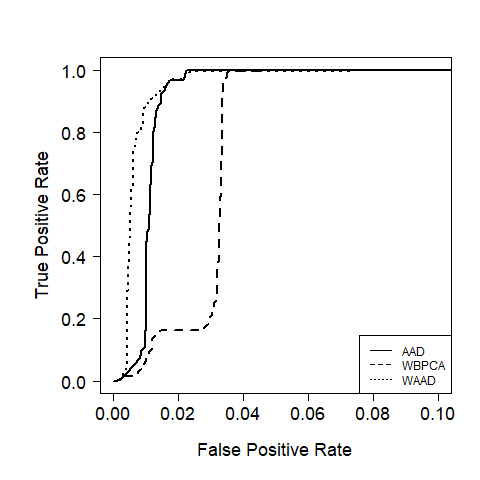}}
\caption{ROC curves for all categories of attacks in UNSW-NB15 dataset.}\label{UNSW-NB15-roc}
\end{figure}

\begin{figure}
\begin{center}$
\begin{array}{c}
\includegraphics[scale=0.250]{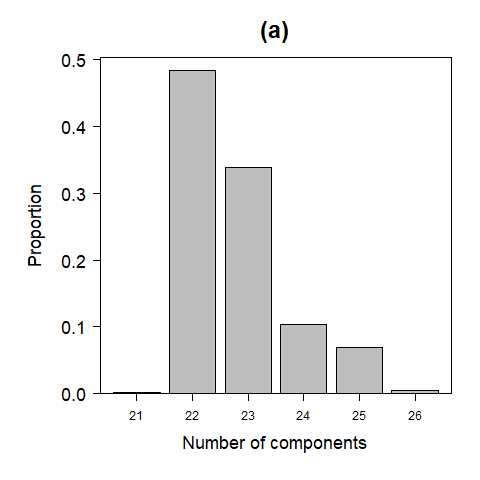}
\includegraphics[scale=0.250]{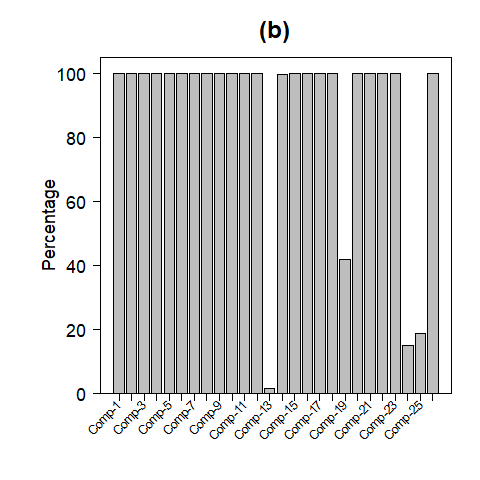}\\
\includegraphics[scale=0.250]{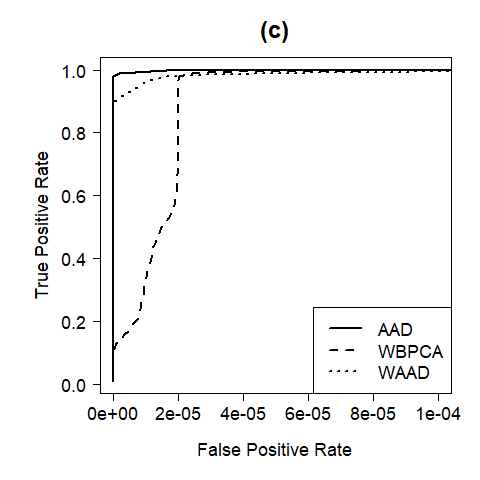}
\end{array}$
\end{center}
\caption{Plots of the UNSW-NB15 dataset. A random sample of size 9900 from training dataset is contaminated with 100 attack connections randomly chosen from all categories of attacks. In (a) is the distribution of the number of components found relevant, in (b) are the components found effected and in (c) are the average ROC curves. The experiment is repeated 1000 times.}\label{UNSW-NB15-all}
\end{figure}

\begin{figure}
\begin{center}$
\begin{array}{c}
\includegraphics[scale=0.250]{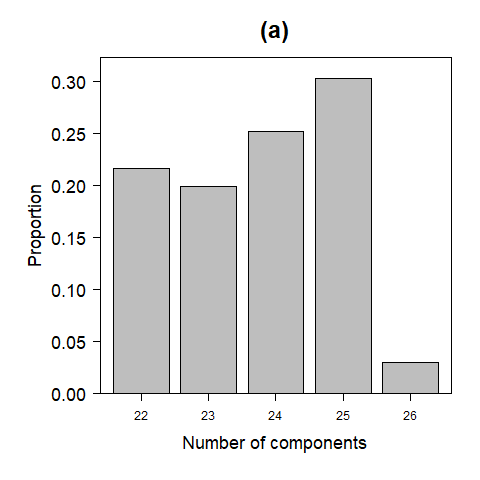}
\includegraphics[scale=0.250]{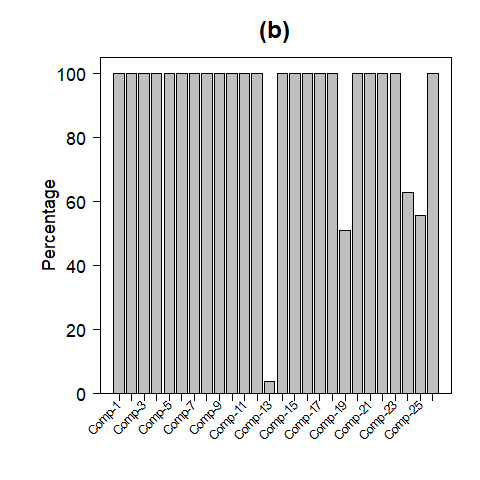}\\
\includegraphics[scale=0.250]{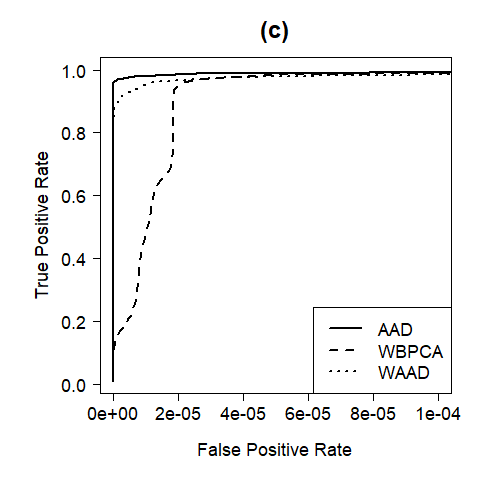}
\end{array}$
\end{center}
\caption{Plots of the UNSW-NB15 dataset. A random sample of size 9900 from training dataset is contaminated with 100 attack connections randomly chosen from `DoS' attacks. In (a) is the distribution of the number of components found relevant, in (b) are the components found effected and in (c) are the average ROC curves. The experiment is repeated 1000 times.}\label{UNSW-NB15-DoS}
\end{figure}

\begin{figure}
\begin{center}$
\begin{array}{c}
\includegraphics[scale=0.250]{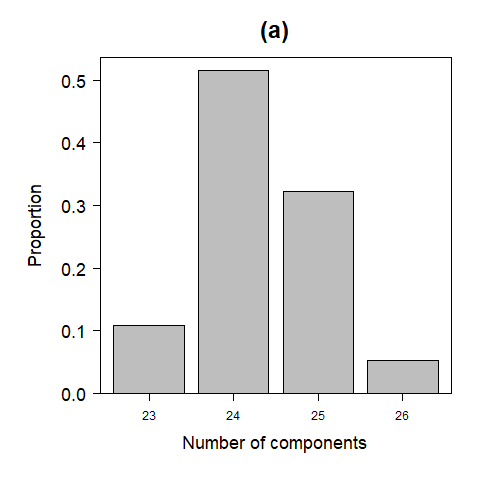}
\includegraphics[scale=0.250]{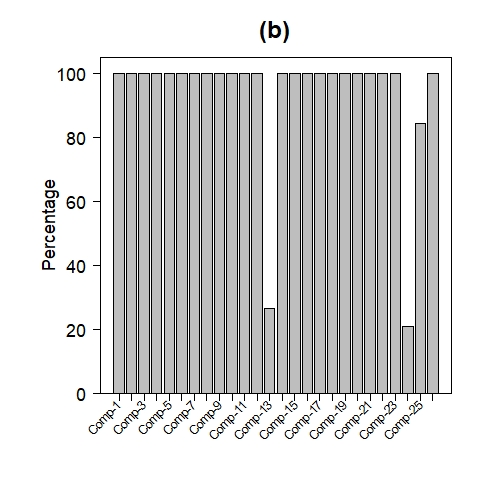}\\
\includegraphics[scale=0.250]{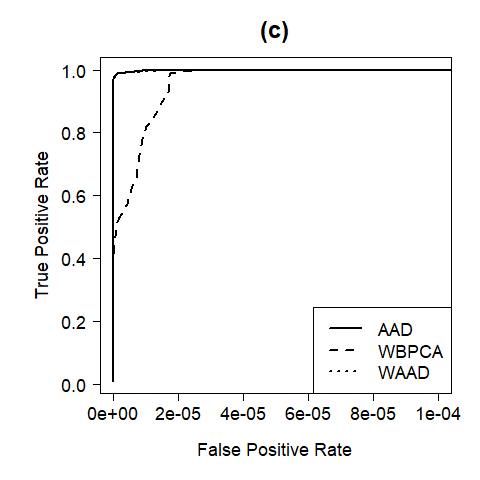}
\end{array}$
\end{center}
\caption{Plots of the UNSW-NB15 dataset. A random sample of size 9900 from training dataset is contaminated with 100 attack connections randomly chosen from `Fuzzers' attacks. In (a) is the distribution of the number of components found relevant, in (b) are the components found effected and in (c) are the average ROC curves. The experiment is repeated 1000 times.}\label{UNSW-NB15-Fuzzers}
\end{figure}

\begin{figure}
\begin{center}$
\begin{array}{c}
\includegraphics[scale=0.250]{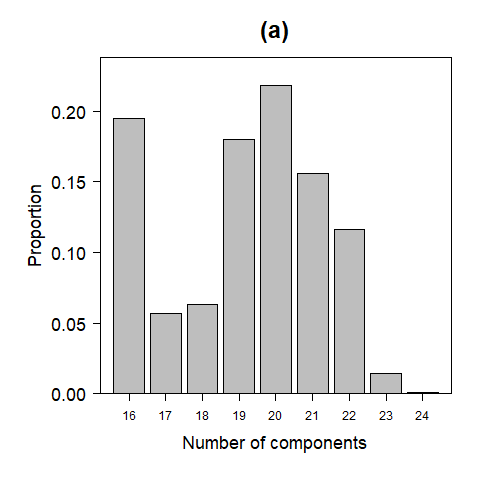}
\includegraphics[scale=0.250]{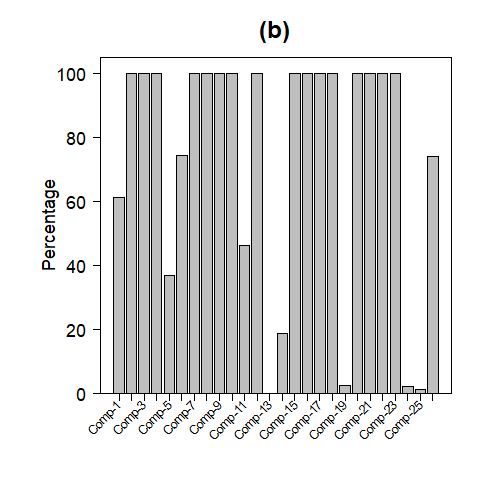}\\
\includegraphics[scale=0.250]{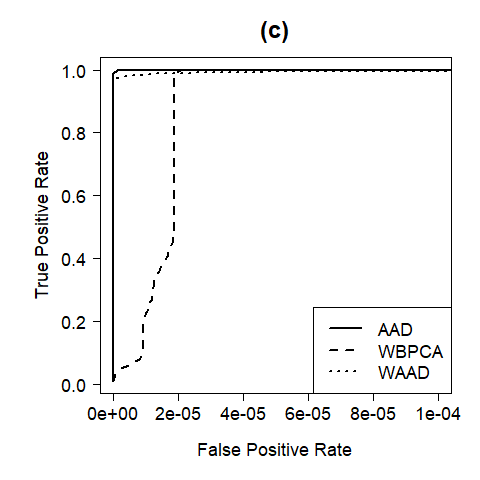}
\end{array}$
\end{center}
\caption{Plots of the UNSW-NB15 dataset. A random sample of size 9900 from training dataset is contaminated with 100 attack connections randomly chosen from `Generic' attacks. In (a) is the distribution of the number of components found relevant, in (b) are the components found effected and in (c) are the average ROC curves. The experiment is repeated 1000 times.}\label{UNSW-NB15-Generic}
\end{figure}

We used the AAD, WAAD and WBPCA as detailed in Section \ref{subsection2.1} to try to detect and identify the attack connections. For the WBPCA, we calculated the eigenvalues using the correlation scale and retained components using the Kaiser's rule also referred to as ``eigenvalue greater than one rule'' \citep{kaiser1960application}. We projected the entire test data onto the eigenspace obtained using the training data and found that some of the $s_j^u$'s are strongly affected (for example, see the largest four $s_j^u$'s with indices 6, 11, 21, 26 in Figure \ref{UNSW-NB15-delta}). The plots for the score statistic $t_i^f$ based on the methods AAD, WAAD and WBPCA are portrayed in Figure \ref{UNSW-NB15-f1}. To see the patterns in the test set of data in comparison to the training set of data, we also plot our training data together with the test set of data, where the training set of the data is shaded in light-blue in each plot of the figure. Overall, the methods are able to detect and identify attacks equally well (AAD, WAAD and WBPCA detect and identify, respectively, 99.836\%, 99.876\% and 99.830\% of the anomalous connections). Figure \ref{UNSW-NB15-f1} reveals that the normal activities observations of the test data observed during the attack periods also exhibit some peculiar patterns that indicate anomalous connections (in that the corresponding $t_i^f$ exceeds the threshold $\theta^{(1-.0001)}$). The normal activities data observed during attack periods also show some anomalous block-like patterns that cannot be seen in the period when clean data were observed. The false alarm rate is high for all methods due to these anomalous patterns of normal activities data happening during the attack periods (see Figure \ref{UNSW-NB15-f1} and Figure \ref{UNSW-NB15-roc}). Although no concrete conclusion could be drawn, it could be that the normal activities are disturbed by the attacks or that some anomalous connections are mislabeled as normal. We also calculated the score statistic $t_i^f$ based on the four most affected dimensions (see the largest $s_j^u$'s with indices 6, 11, 21, and 26 in Figure \ref{UNSW-NB15-delta}) and plot the results in Figure \ref{UNSW-NB15-f1}(b). Considering only these four components in the final score $t_i^f$ was almost as good in terms of detection and identification of anomalous connections and other anomalous patterns, since it further pushed the block like anomalous structures across the threshold of normal activities data.

Often the attacks do not happen all at once, in which case it may be more natural to try to detect and identify a smaller number of anomalous connections in a large batch of normal connections. We therefore contaminated a large sample of size 9900 randomly drawn from the training data with 100 connections randomly drawn from the set of anomalous connections ($m=9900+100$) and used all three methods (AAD, WAAD and WBPCA) to detect and identify the 100 attack connections. We repeated the experiment 1000 times and show the results in Figure \ref{UNSW-NB15-all}. The results indicated that 22 components out of 26 were affected by the anomalies (see Figure \ref{UNSW-NB15-all} (a)) and they were not necessarily only among the first few components (see Figure \ref{UNSW-NB15-all} (b)). Taking only the affected components into account made the NIDS slightly more effective both in terms of detection rate and false positive rate (see Figure \ref{UNSW-NB15-all} (c)). We also repeated the experiments separately for each type of attack and show the results for `DoS', `Fuzzers' and `Generic' attacks , respectively, in Figures \ref{UNSW-NB15-DoS}, \ref{UNSW-NB15-Fuzzers} and \ref{UNSW-NB15-Generic}. The performance of the proposed methods was found to be superior in detecting individual types of attacks. Note that that most of the attacks affected the same components both in combination and separately, which may be due to similar patterns across different types of attacks as exhibited in Figure \ref{UNSW-NB15-f1}.

\subsection{KDD'99 dataset}
\label{subsection4.2}
The KDD'99 dataset\footnote{\url{http://kdd.ics.uci.edu/databases/kddcup99/kddcup99.html}} is perhaps the most well-known and wildly used reference network dataset for the evaluation of NIDS \citep{tavallaee2009detailed}. It was released by MIT Lincoln Lab for a 1998 DARPA evaluation program that was aimed to survey and evaluate research in intrusion detection.

The training dataset consists of 4,898,431 connections, which was prepared from 4 gigabytes of compressed raw tcpdump data of 7 weeks of network traffic by \cite{stolfo2000cost}. It contains 972,781 normal connections and 3,925,650 anomalous connections. The anomalous connections are due to 23 types of attacks listed in Table \ref{atttypes} and each of which falls into one of 4 categories: denial-of-service, unauthorized access from a remote machine, unauthorized access to local superuser privileges by a local unprivileged user, and surveillance and probing \citep{lee2000framework}. Some of these 23 types of attacks are rare; for example, `spy' happens two times and `perl' happens three times in the whole dataset. Others happen frequently in the dataset; for example, there are 2,807,886 `smurf' connections.

\begin{table}{!tb}
\caption{Attack types in KDD'99 dataset.\label{atttypes}}
\centering
\begin{tabular}{rr}
  \hline
Attack type & Frequency \\
  \hline
back & 2203 \\
  buffer\_overflow &  30 \\
  ftp\_write &   8 \\
  guess\_passwd &  53 \\
  imap. &  12 \\
  ipsweep & 12481 \\
  land. &  21 \\
  loadmodule &   9 \\
  multihop &   7 \\
  neptune & 1072017 \\
  nmap & 2316 \\
  normal & 972781 \\
  perl &   3 \\
  phf &   4 \\
  pod & 264 \\
  portsweep & 10413 \\
  rootkit &  10 \\
  satan & 15892 \\
  smurf & 2807886 \\
  spy &   2 \\
  teardrop & 979 \\
  warezclient & 1020 \\
  warezmaster &  20 \\
   \hline
\end{tabular}
\end{table}

Similarly, around two million connection records were obtained separately from the two weeks of test data. However, following \cite{wang2006identifying}, we used only the training set and drop the original test data completely.

The dataset has 41 features excluding the one that represents class labels \citep{lee2000framework}. Among these 41 features, 34 are numeric and 7 are
nominal. We include only 23 numeric features in our experiments and exclude all nominal features and 11 numeric features (`wrong\_fragment', `urgent', `hot', `num\_failed\_logins', `num\_compromised', `root\_shell', `su\_attempted', `num\_root' and `num\_file\_creations') since these feature have a small number of non-zero values in the training set of data (the values for `wrong\_fragment' and `num\_outbound\_cmds' are zero for all connections).

\begin{figure}[!t]
\centering
\subfloat{\includegraphics[scale=0.300]{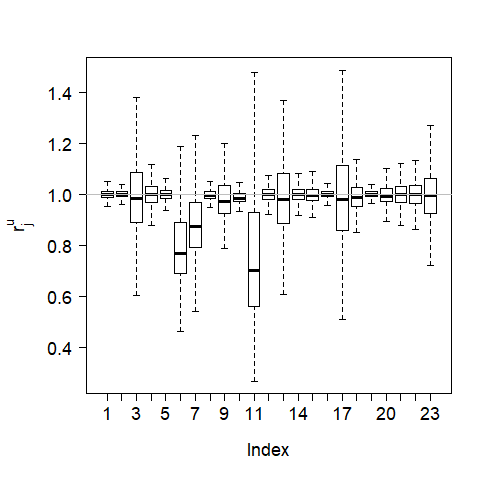}}\\
\caption{Plot of the values of $r_j^u$'s obtained using 5000 bootstrap samples of size 10000 from the training set of the KDD'99 dataset. The horizontal gray line indicates the expected value of $r_j^u$ in the absence of anomalous activities. }\label{KDD99-outliers}
\end{figure}

\begin{figure}[!t]
\centering
\subfloat{\includegraphics[scale=0.300]{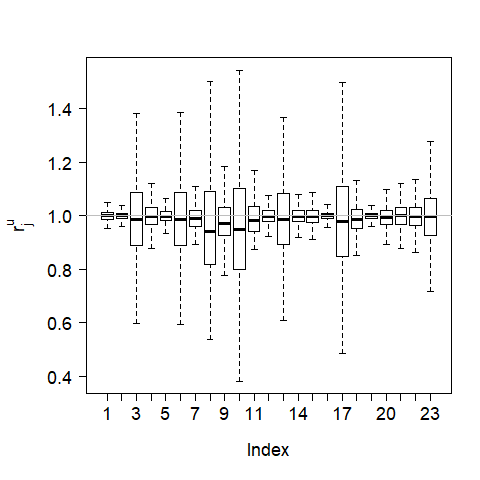}}\\
\caption{Plot of the values of $r_j^u$'s obtained using 5000 bootstrap samples of size 10000 from the training set of the KDD'99 dataset after removal of eight unusual observations. The horizontal gray line indicates the expected value of $r_j^u$ in the absence of anomalous activities. }\label{KDD99-wout}
\end{figure}

We used $n=972,781$ normal connections data for training the PCA model. As in the previous application, we constructed the distributions of $r_j^u$'s based on 5,000 bootstrap samples of size 10,000 (see Figure \ref{KDD99-outliers}). The distributions of some of the $r_j^u$'s were centered away from one (see the box-plots for $r_6^u$, $r_7^u$ and $r_{11}^u$), indicating the possible presence of some unusual observations of the variables that have large coefficients for PC-6, PC-7 and PC-11. The PC-6, PC-7 and PC-11 all had large coefficients for `src\_bytes' and `dst\_bytes' (PC-6 had loadings of -0.51 and -0.58, PC-7 had loadings of -0.46 and -0.39, and PC-11 had loadings of -0.70 and 0.70, respectively for `src\_bytes' and `dst\_bytes'. Further  inspection revealed eight unusually large values of the `src\_bytes' and eight unusually large values of the `dst\_bytes'. The removal of these 16 unusual observations re-centered the distributions of $r_6^u$, $r_7^u$ and $r_{11}^u$ closer to one (see Figure \ref{KDD99-wout}).

\begin{figure}[!t]
\begin{center}$
\begin{array}{c}
\includegraphics[scale=0.250]{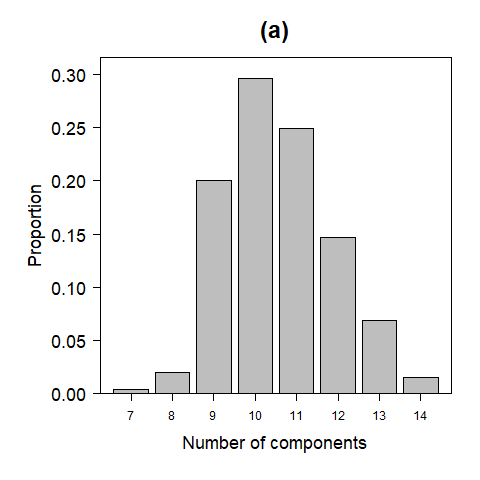}
\includegraphics[scale=0.250]{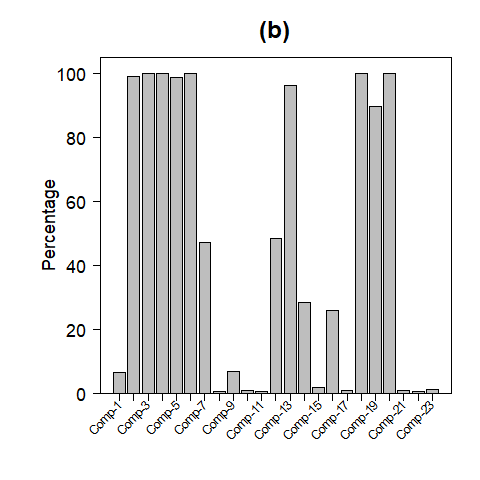}\\
\includegraphics[scale=0.250]{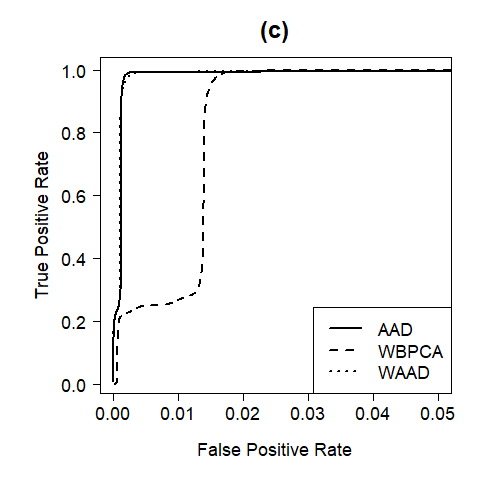}
\end{array}$
\end{center}
\caption{Plots of the KDD'99 dataset. A random sample of size 9900 from normal activities dataset is contaminated with 100 attack connections randomly chosen from all categories of attacks. In (a) is the distribution of the number of components found relevant, in (b) are the components found effected and in (c) are the average ROC curves. The experiment is repeated 1000 times.}\label{KDD99-full}
\end{figure}

\begin{figure}[!t]
\begin{center}$
\begin{array}{c}
\includegraphics[scale=0.250]{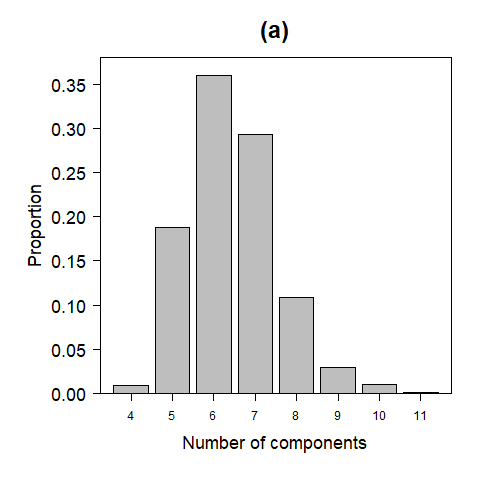}
\includegraphics[scale=0.250]{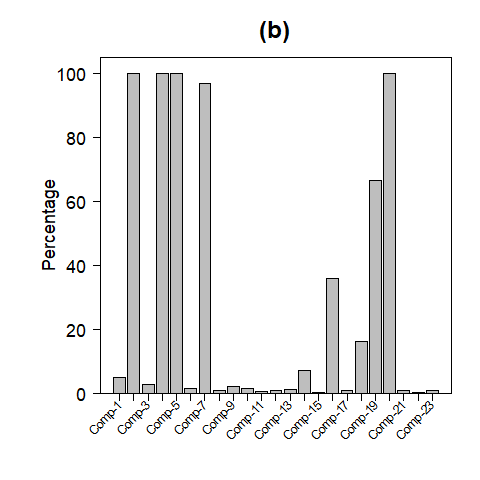}\\
\includegraphics[scale=0.250]{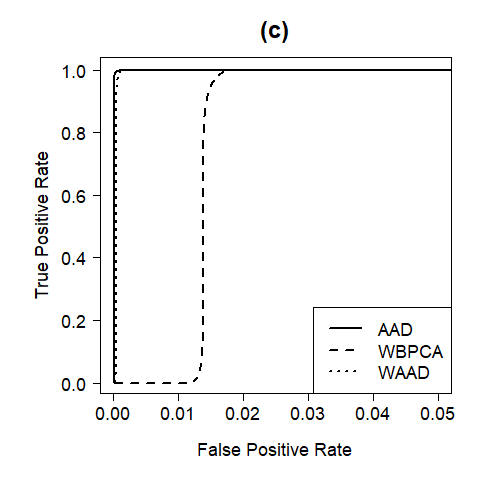}
\end{array}$
\end{center}
\caption{Plots of the KDD'99 dataset. A random sample of size 9900 from normal activities dataset is contaminated with 100 attack connections randomly chosen from `smurf' attacks. In (a) is the distribution of the number of components found relevant, in (b) are the components found effected and in (c) are the average ROC curves. The experiment is repeated 1000 times.}\label{KDD99-smurf}
\end{figure}

\begin{figure}[!t]
\begin{center}$
\begin{array}{c}
\includegraphics[scale=0.250]{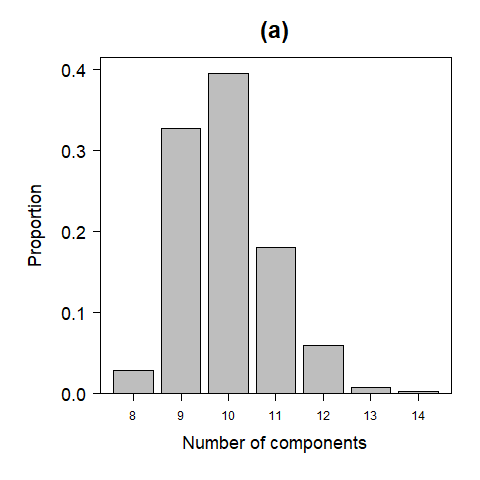}
\includegraphics[scale=0.250]{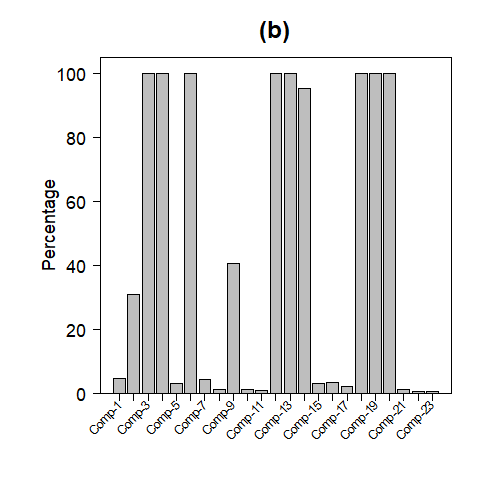}\\
\includegraphics[scale=0.250]{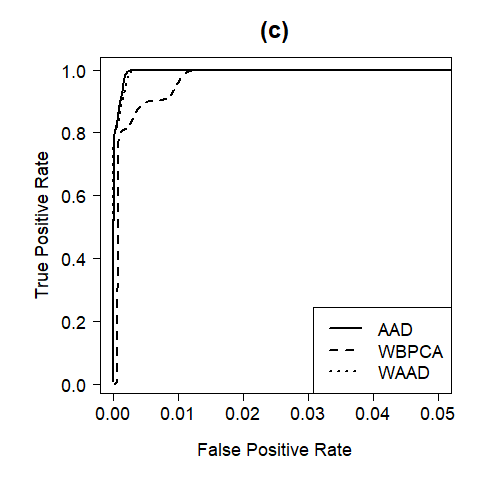}
\end{array}$
\end{center}
\caption{Plots of the KDD'99 dataset. A random sample of size 9900 from normal activities dataset is contaminated with 100 attack connections randomly chosen from `neptune' attacks. In (a) is the distribution of the number of components found relevant, in (b) are the components found effected and in (c) are the average ROC curves. The experiment is repeated 1000 times.}\label{KDD99-neptune}
\end{figure}

\begin{figure}[!t]
\begin{center}$
\begin{array}{c}
\includegraphics[scale=0.250]{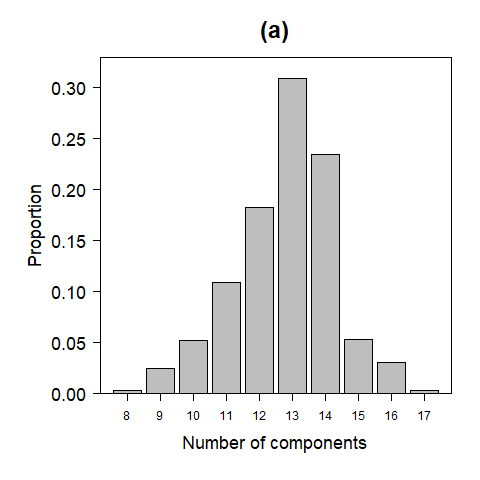}
\includegraphics[scale=0.250]{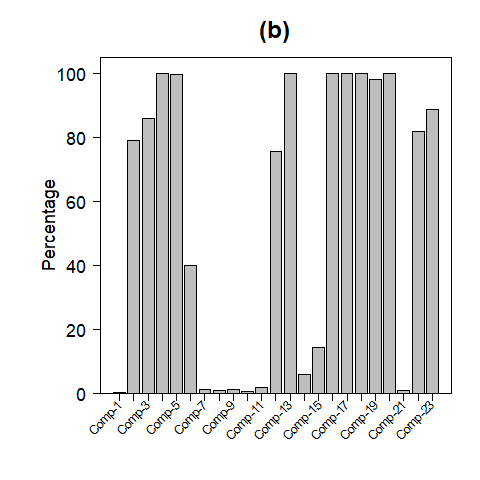}\\
\includegraphics[scale=0.250]{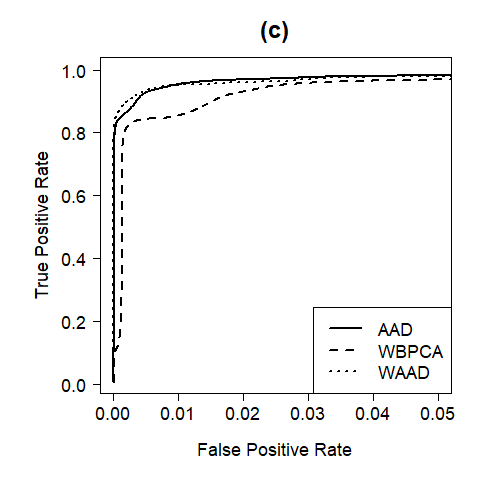}
\end{array}$
\end{center}
\caption{Plots of the KDD'99 dataset. A random sample of size 9900 from normal activities dataset is contaminated with 100 attack connections randomly chosen from `satan' attacks. In (a) is the distribution of the number of components found relevant, in (b) are the components found effected and in (c) are the average ROC curves. The experiment is repeated 1000 times.}\label{KDD99-satan}
\end{figure}

\begin{figure}[!t]
\begin{center}$
\begin{array}{c}
\includegraphics[scale=0.250]{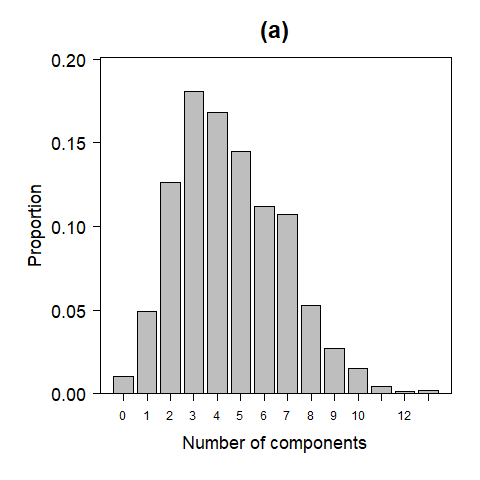}
\includegraphics[scale=0.250]{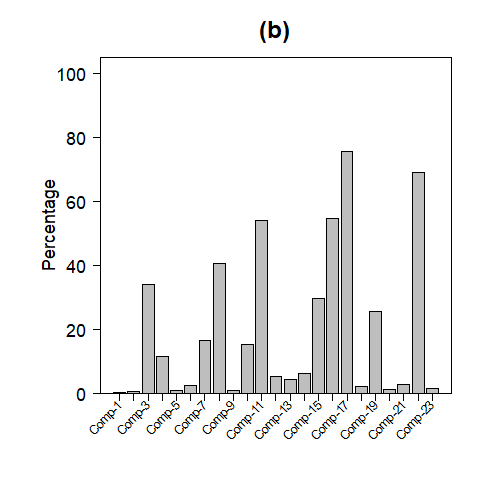}\\
\includegraphics[scale=0.250]{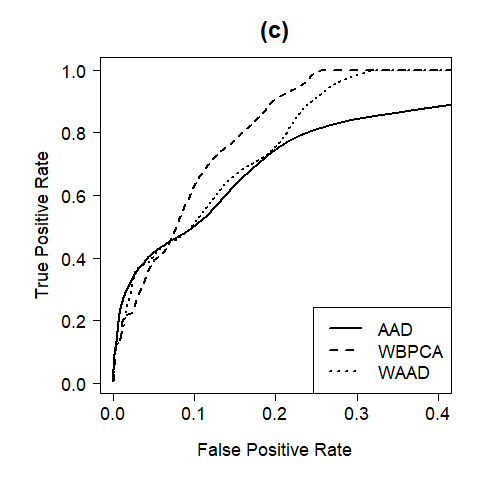}
\end{array}$
\end{center}
\caption{Plots of the KDD'99 dataset. A random sample of size 9900 from normal activities dataset is contaminated with 100 attack connections randomly chosen from all less frequent attacks. In (a) is the distribution of the number of components found relevant, in (b) are the components found effected and in (c) are the average ROC curves. The experiment is repeated 1000 times.}\label{KDD99-rare}
\end{figure}

For the test data, we contaminated a large sample of size 9900 randomly drawn from normal activities data with 100 anomalous connections randomly drawn from the set of anomalous connections ($m=9900+100$) and used the AAD, WAAD and WBPCA to detect and identify the 100 attack connections. We repeated the experiment 1000 times and show the results in Figure \ref{KDD99-full}. The results indicated that there were indeed a few components out of 32 that were affected by the anomalies and they were not necessarily only among the first few components (see Figure \ref{KDD99-full} (a) and (b)). Taking only the affected components into account made the NIDS much more effective both in terms of detection rate and false positive rate (see Figure \ref{KDD99-full} (c)). We also repeated the experiments separately for the three most frequent attacks, namely `smurf', `neptune' and `satan', and show the results, respectively, in Figures \ref{KDD99-smurf}, \ref{KDD99-neptune} and \ref{KDD99-satan}. The performance of the proposed methods was found to be superior in detecting individual type of attacks.

As mentioned earlier, some of the attacks were rare and the probability for them to be included in a random sample of sized 100 was negligible. Therefore, we repeated the experiment with all attacks which occurred less than 1,000 times. The results are presented in Figure \ref{KDD99-rare}. The number of affected components was less stable due to the mixture of varieties of rare attacks each of which altered a varying number of components (see Figure \ref{KDD99-rare} (a) and (b)) compared to the case when individual type of attacks were confronted. The detection rate was low and comparable for all methods. This indicates that the proposed methods AAD and WAAD perform similarly to WBPCA when varieties of attacks are introduced at the same time and improve the detection rate when normal connections are contaminated with a single type of attack at a time.

\section{Discussion and conclusion}
\label{section5}
The use of PCA for network anomaly detection has been criticized due to its limited effectiveness in detecting and identifying anomalous activities \citep{ringberg2007sensitivity}.
One reason for this is that the anomaly scores are calculated by aggregation across the major principal components (the first few components that explain most of the variation in the data), which often includes components that are relatively unaffected by the anomalies, or the alterations caused by anomalies along these components are less pronounced because of greater variation along these components. Moreover, the existing methods exclude some of the minor components from the aggregation that could be more strongly affected, or the alterations caused by anomalies along these components are more pronounced because of lower variation along these components. In this paper, we highlight this shortcoming of previous studies and propose to calculate scores by aggregating across only those components that are affected by the anomalies or give more weight to the components that are more affected.

To support our findings, we have conducted a simulation study which showed that the identification rate could be considerably increased if the scores were aggregated only across affected components while leaving out the unaffected components. Anomaly identification was higher when the most affected components were weighted more highly. Through the analysis of real datasets we showed that the attacks created shifts only along the direction of a few components (a mix of principal and minor components) and that most of the components are relatively unaffected, so when excluded from the calculation of anomaly scores, the detection rate improves.

Although, our simulation study is limited to the multivariate normal distribution, the applications to real data indicated that that the proposed methods can perform well even if the data is non-normal. However, further simulations---simulations from heavy-tailed and skewed distributions---would strengthen this assertion.

\label{Bibliography}
\bibliographystyle{IEEEtran}
\bibliography{Bibliography}

%








\end{document}